\documentclass[onecolumn]{aastex61}
\usepackage{color}
\usepackage{graphicx}
\usepackage{grffile}
\usepackage{bm}
\usepackage{xparse}
\usepackage{enumerate}
\usepackage{verbatim}
\usepackage{natbib}
\usepackage{longtable}
\usepackage{threeparttablex}

\usepackage{float}
\usepackage[utf8]{inputenc}
\DeclareUnicodeCharacter{FB01}{fi}

\usepackage[
  separate-uncertainty = true,
  multi-part-units = repeat
]{siunitx}

\begin{document}

\title{CORONAL DIMMINGS ASSOCIATED WITH CORONAL MASS EJECTIONS ON THE SOLAR LIMB}

\correspondingauthor{Galina Chikunova}
\email{galina.chikunova@skoltech.ru}

\author{Galina Chikunova}
\affiliation{Skolkovo Institute of Science and Technology,
Bolshoy Boulevard 30, bld. 1, Moscow 121205, Russia}
\author{Karin Dissauer}
\affiliation{NorthWest Research Associates, 3380 Mitchell Lane, Boulder 80301, CO, USA}
\affiliation{Institute of Physics, University of Graz, Universit{\"a}tsplatz 5, 8010 Graz, Austria}
\author{Tatiana Podladchikova}
\affiliation{Skolkovo Institute of Science and Technology,
Bolshoy Boulevard 30, bld. 1, Moscow 121205, Russia}
\author{Astrid M. Veronig}
\affiliation{Institute of Physics, University of Graz, Universit{\"a}tsplatz 5, 8010 Graz, Austria}
\affiliation{Kanzelh\"ohe Observatory for Solar and Environmental Research, University of Graz, Kanzelh{\"o}he 19, 9521 Treffen, Austria}

\begin{abstract}
We present a statistical analysis of 43 coronal dimming events, associated with Earth-directed CMEs that occurred during the period of quasi-quadrature of the SDO and STEREO satellites. We studied coronal dimmings that were observed above the limb by STEREO/EUVI and compared their properties with the mass and speed of the associated CMEs. The unique position of satellites allowed us to compare our findings with the results from \cite{dissauer2018statistics,dissauer2019statistics}, who studied the same events observed against the solar disk by SDO/AIA. Such statistics is done for the first time and confirms the relation of coronal dimmings and CME parameters for the off-limb viewpoint.
The observations of dimming regions from different lines-of-sight reveal a similar decrease in the total EUV intensity ($c=0.60\pm0.14$). We find that the (projected) dimming areas are typically larger for off-limb observations (mean value of $1.24\pm1.23\times10^{11}$ km$^2$ against $3.51\pm0.71\times10^{10}$ km$^2$ for on-disk), with a correlation of $c=0.63\pm0.10$. This systematic difference can be explained by the (weaker) contributions to the dimming regions higher up in the corona, that cannot be detected in the on-disk observations. The off-limb dimming areas and brightnesses show very strong correlations with the CME mass ($c=0.82\pm0.06$ and $c=0.75\pm0.08$), whereas the dimming area and brightness change rate correlate with the CME speed ($c\sim0.6$). Our findings suggest that coronal dimmings have the potential to provide early estimates of mass and speed of Earth-directed CMEs, relevant for space weather forecasts, for satellite locations both at L1 and L5.

\end{abstract}

\section{Introduction} \label{sec:intro}
Coronal Mass Ejections (CMEs) are the most energetic and powerful eruptive phenomena on the Sun and also the main drivers of space weather effects. Huge amounts of magnetized plasma are expelled from the Sun to interplanetary space with speeds in the range of some 100 km s$^{-1}$ up to $>$3000km s$^{-1}$ \mbox{\citep{2009EM&P..104..295G,Webb2012}} and may interact with the Earth magnetosphere and atmosphere to produce geomagnetic storms and affect human environment and technologies \citep{gosling1993solar,lanzerotti2017space,knipp2018little}.

Ironically, Earth-directed CMEs are most difficult to measure using observations from instruments along the Sun-Earth line, due to strong projection effects \citep{burkepile2004role}. In these cases, we mostly observe the CME's expansion motion and not its propagation towards Earth. In addition, their initiation and early acceleration are also difficult to observe as these regions are blocked by the coronagraph's occulter disk. The operation of NASA's STEREO mission \citep{2008SSRv..136....5K} with its unique capability of imaging the inner heliosphere from two different vantage points away from the Sun-Earth line has greatly improved our understanding of CME structure and evolution \citep{Aschwanden2009,rouillard2011relating,thernisien2011cme}. Depending on the position on their orbits around the Sun, the STEREO satellites provide us with the capability to observe Earth-directed CMEs from the side, including observations of their initiation and early evolution above the limb in Extreme-Ultraviolet (EUV) and coronagraph observations by the EUVI and COR instruments.

Coronal dimmings are transient regions of strongly reduced emission in soft X-rays (SXR) \citep{Hudson:1996,Sterling:1997} and EUV \citep{Thompson:1998,1999ApJ...520L.139Z} wavelengths that occur in association with CMEs. In general, they are interpreted as density depletion caused by mass loss during the CME eruption \citep{2000JGR...10527251W}. The loss of mass in coronal dimmings is also evidenced by spectroscopic studies that identified plasma outflows in coronal dimming regions
\citep[e.g.][]{harrison2000spectroscopic,harra2001material,2012ApJ...748..106T,2019ApJ...879...85V}. In addition, Differential Emission Measure (DEM) analysis of coronal dimmings showed that the impulsive decrease of emission in dimming regions stems mostly from a distinct drop in density (up to 70\%), whereas the  variations in plasma temperature are much smaller \citep{lopez2017mass,2018ApJ...857...62V,2019ApJ...879...85V}. Coronal dimmings are also closely associated with large-scale coronal waves initiated by CMEs \citep[e.g.][]{thompson1998soho,dissauer2016projection,veronig2018genesis,podladchikova2019three}. Thus, coronal dimmings contain important information on the early CME evolution, characteristic CME properties as well as associated shock waves.

Different approaches have been developed and used for the extraction of coronal dimmings: thresholding pixel values from base-difference images in \cite{reinard2008coronal} and \cite{bewsher2008relationship}, applying techniques of producing minimum and maximum intensity maps in the NEMO algorithm by \cite{podladchikova2005automated} and its extension by \cite{Attrill_2009}, using the thresholded intersection of running-difference and percentage running-difference images in the Solar Demon algorithm by \cite{refId0}, constructing Lambert projection maps from direct EUV data by the CoDiT software in \cite{krista2012study}, applying a thresholding-based algorithm using logarithmic base ratio images in \cite{dissauer2018detection}.

The automated algorithms allow to process a large number of events and create a statistical basis of coronal dimmings and their associated CMEs. However, there exist only a few statistical studies of coronal dimmings \citep{bewsher2008relationship,reinard2008coronal,aschwanden2016global,mason2016relationship,krista2017statistical}. A thorough methodological approach and comprehensive statistical analysis of the characteristic properties of coronal dimmings, and their relation to the decisive parameters of the associated CMEs and flares has recently been presented in a series of papers by \cite{dissauer2018detection, dissauer2018statistics, dissauer2019statistics}. In these studies 62 dimming events were analyzed using multi-viewpoint observations by the SDO and STEREO satellites in quasi quadrature: coronal dimmings were observed on-disk by SDO/AIA \citep{2012SoPh..275...17L} and the related magnetic fluxes were obtained using SDO/HMI, while the kinematics and the mass of the associated CME was observed in STEREO EUVI and COR data close to the limb, minimizing projection effects. These studies revealed that the CME mass shows high correlations with the dimming area, its total brightness, and magnetic flux, whereas the  maximal speed of the CME is strongly correlated with their corresponding time derivatives (i.e. area growth rate, brightness change rate, and magnetic flux rate) with correlation coefficients in the range of $c \sim 0.6 - 0.7$ \citep{dissauer2018statistics, dissauer2019statistics}.

In this paper, we study the same data set, but now extract the coronal dimming properties and their evolution from STEREO/EUVI, which observes the dimmings above the limb. This approach allows us for the first time to compare the results of on-disk and off-limb observations of coronal dimmings, and also to study how the viewing position affects the derived properties.

\begin{figure}[H]
\centering
\includegraphics[width=0.9\columnwidth]{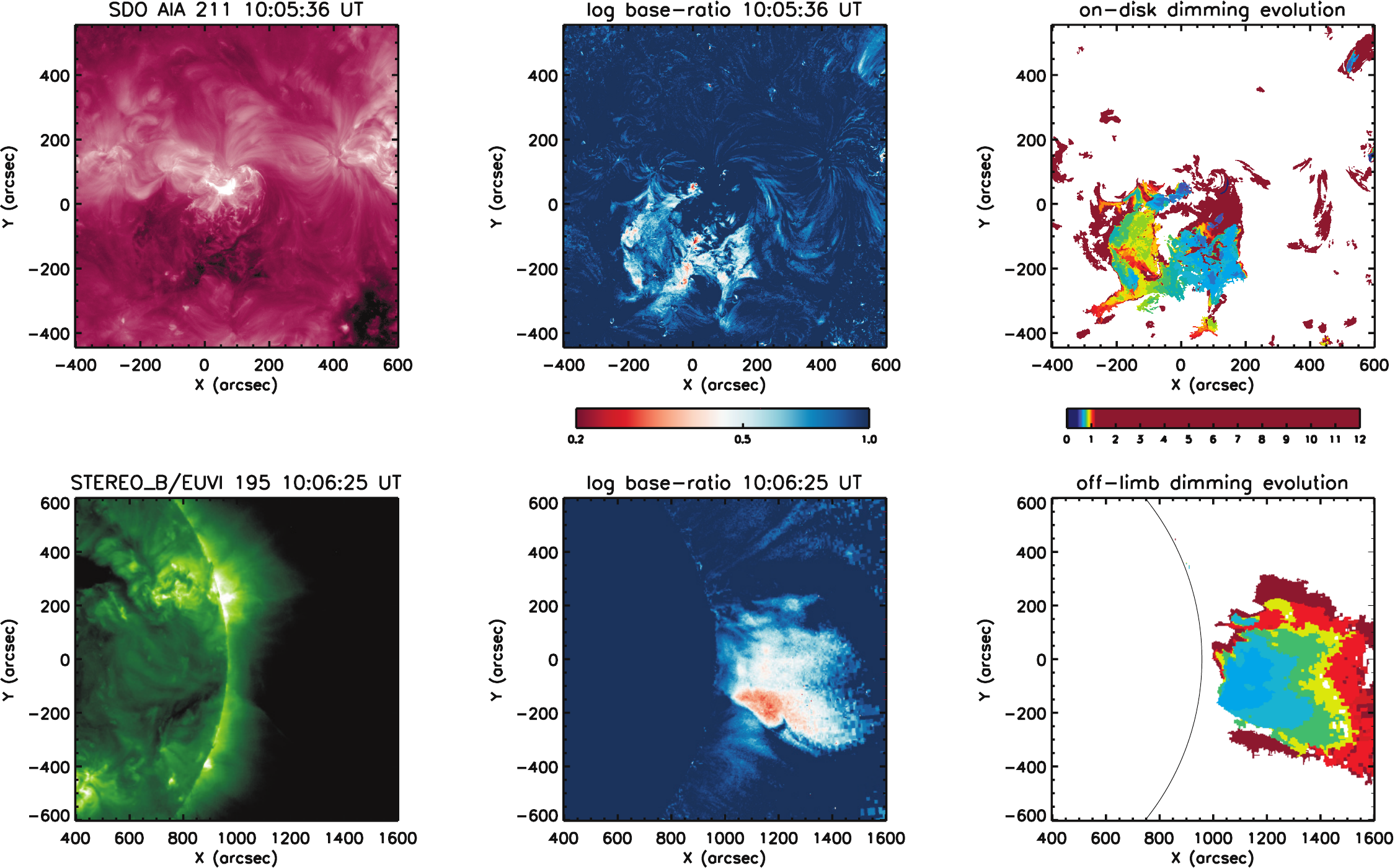}
\caption{Illustration of the dimming evolution from two different viewpoints for the event of 2011 October 1. The top panels represent on-disk dimming observations: SDO AIA 211{\AA} filtegram (left), logarithmic base-ratio map (middle), timing map (right), that marks when each dimming pixel was detected for the first time (presented in hours after the flare onset). The bottom panels show the same for the off-limb dimming observations by STEREO-B/EUVI.}
\label{fig:on-off}
\end{figure}

For illustration, we show Figure \ref{fig:on-off} as an example of the different appearance of the dimming event that occurred on 2011 October 1, observed on-disk by SDO AIA \citep{dissauer2019statistics} and off-limb by the STEREO-B satellite. We show the direct SDO AIA 211{\AA} (top) and the STEREO-B/EUVI 195{\AA} (bottom) images close to the time of maximal extent of the dimming, the corresponding logarithmic base-ratio images and the timing maps. The colors in the timing maps encode for each dimming pixel the time of its first detection.

\section{Data and Data Reduction}
In this study, we use data from NASA's Solar Terrestrial Relations Observatory \cite[STEREO;][]{2008SSRv..136....5K}. The Extreme-Ultraviolet Imagers \citep[EUVI;][]{2004SPIE.5171..111W} of the STEREO/SECCHI instrument suite provide images of the Sun with a pixel scale of 1.6 arcsec through four EUV filters.  We used filtergrams of the 195 {\AA}  passband with a cadence of 5 minutes (except for individual events where it can vary from 1.5 min to 10 min). According to \cite{dissauer2018detection, refId0}, coronal dimmings are best observed in wavelengths sensitive to quiet Sun coronal temperatures (e.g. 195 {\AA}, 171 {\AA}, 211 {\AA}). 195 {\AA} has the highest cadence for STEREO/EUVI and therefore we use this wavelength.

We study the time range between May 2010 and September 2012 when SDO was in quasi-quadrature with the STEREO-A and STEREO-B satellites, increasing their separation angle to the Sun-Earth line during this period from $\pm75^{\circ}$  to $\pm125^{\circ}$. We focus on 43 events in which coronal dimmings and their associated Earth-directed CMEs were observed off-limb by at least one of the two STEREO satellites. 39 of these events overlap with the event set from \cite{dissauer2018statistics,dissauer2019statistics} who studied the same dimmings as observed on-disk by the Atmospheric Imaging Assembly (AIA) instrument on-board NASA's Solar Dynamics Observatory (SDO) satellite \citep{2012SoPh..275...17L}. Four events were excluded in the \cite{dissauer2018statistics,dissauer2019statistics} studies, due to filament material that partially obscured the on-disk dimming detection. This was not the case for the STEREO off-limb observations, and thus they are included in the present work. However, for 23 events of the overall data set of 62 events, the off-limb dimming extraction was not possible due to only partial visibility of the dimmings by the STEREO satellites or the activity of multiple active regions at the same time. One event (2012 July 2) was excluded from the statistics due to its complexity (disappearance of trans-equatorial loops) revealed during the comparative analysis from the two viewpoints.

The kinematical CME parameters used for comparison with the derived dimming properties are from the STEREO/EUVI and COR measurements in \cite{dissauer2019statistics}. The fastest event in the data set occurred on 2012 March 6 with a maximal velocity of $\sim$3700 km s$^{-1}$ and mass of $1.83\times 10^{16}$ g. For the other events the maximal velocities of the CMEs vary from 370 km s$^{-1}$ to 2000 km s$^{-1}$, and the values of CME masses lie between $2.03\times 10^{14}$ g and $1.78\times 10^{16}$ g.

For each event we analyze a series of STEREO EUVI images, which begins 30 minutes before the associated flare and lasts for 12 hours in total.
All EUVI images of the data set were checked for the exposure time, prepared with standard SolarSoft routines (\texttt{secchi\_prep.pro}) and corrected for differential rotation (\texttt{drot\_map.pro}). The images were cut automatically by half (either East or West hemisphere), depending on which satellite observed the event better, STEREO-A or STEREO-B.

\section{Methods and Analysis}
In this paper, we study the evolution of off-limb coronal dimmings and their corresponding CMEs. We derive characteristic dimming parameters, such as its area, brightness and duration, using observations, where the coronal dimmings and the associated CME are observed off-limb by STEREO. The dimming segmentation algorithm is demonstrated for two example events that occurred on 2011 October 1 (event no. 19, impulsive M1.2 flare associated with an EUV wave) and on 2012 March 6 (event no. 29, impulsive X5.4 flare associated with an EUV wave and a halo CME).

\subsection{Dimming extraction algorithm}

We identify the area and the brightness of the dimming regions observed off-limb by applying an automated detection technique based on thresholding and region growing. Usually, base-difference (BD) images are used for the recognition of dimmings \citep[e.g. the algorithms in][]{reinard2008coronal,Attrill_2009}, which show absolute changes in intensity. However, some regions of coronal dimmings represent only a slight decrease in intensity or may develop in the upper corona, where the density is lower. Therefore, such changes are difficult to identify by its absolute value compared to other differences in the coronal intensity.

In \cite{dissauer2018detection} the advantage of using logarithmic base-ratio (LBR) images has been discussed. This form of preprocessing the data is based on dividing each frame of the series by a pre-event image. Thus, these images reflect relative changes in logarithmic intensity, which for the off-limb dimmings also allow us to detect regions higher up in the corona. It is worth mentioning that the application of the logarithm on the ratio images is identical to the difference of the logarithmic images. However, the usage of the ratio also leads to more intensity fluctuations, especially in regions higher up in the corona. Thus, we combine both types of data preprocessing to reduce this noise in the extraction of the dimming regions, while still being sensitive to detect also smaller intensity changes due to dimmings. We use a region-growing algorithm \citep{podladchikova2005automated}, which examines neighboring pixels of initial seed points and determines whether the pixel neighbors should be added to the region or not. It is not possible to define the seed pixels in LBR data due to the noise in the upper corona. Therefore, we combine two types of data preprocessing: we use base-difference images for identification of the darkest pixels, that will later form the seed pixels, applied to the regions in the corresponding areas in logarithmic base-ratio images. Since the seed pixels are generally not directly connected to the fluctuations identified higher up in the corona, this approach provides a robust means to detect and segment dimming regions above the limb for both strong and weak dimmings.

First, we threshold base-difference data by -1.0 DN and extract 30\% of the darkest pixels from it. In this way, the number of pixels with an intensity drop of more than -1.0 DN form the array of seed pixels for the region-growing algorithm (see Fig.~\ref{fig:detection example1}d). We use the darkest pixels in base-difference data because this criterion is independent of a fixed threshold and is therefore not affected by the changes in the mean intensity in the overall corona over time.

\begin{figure}[H]
\centering
\includegraphics[width=0.5\columnwidth]{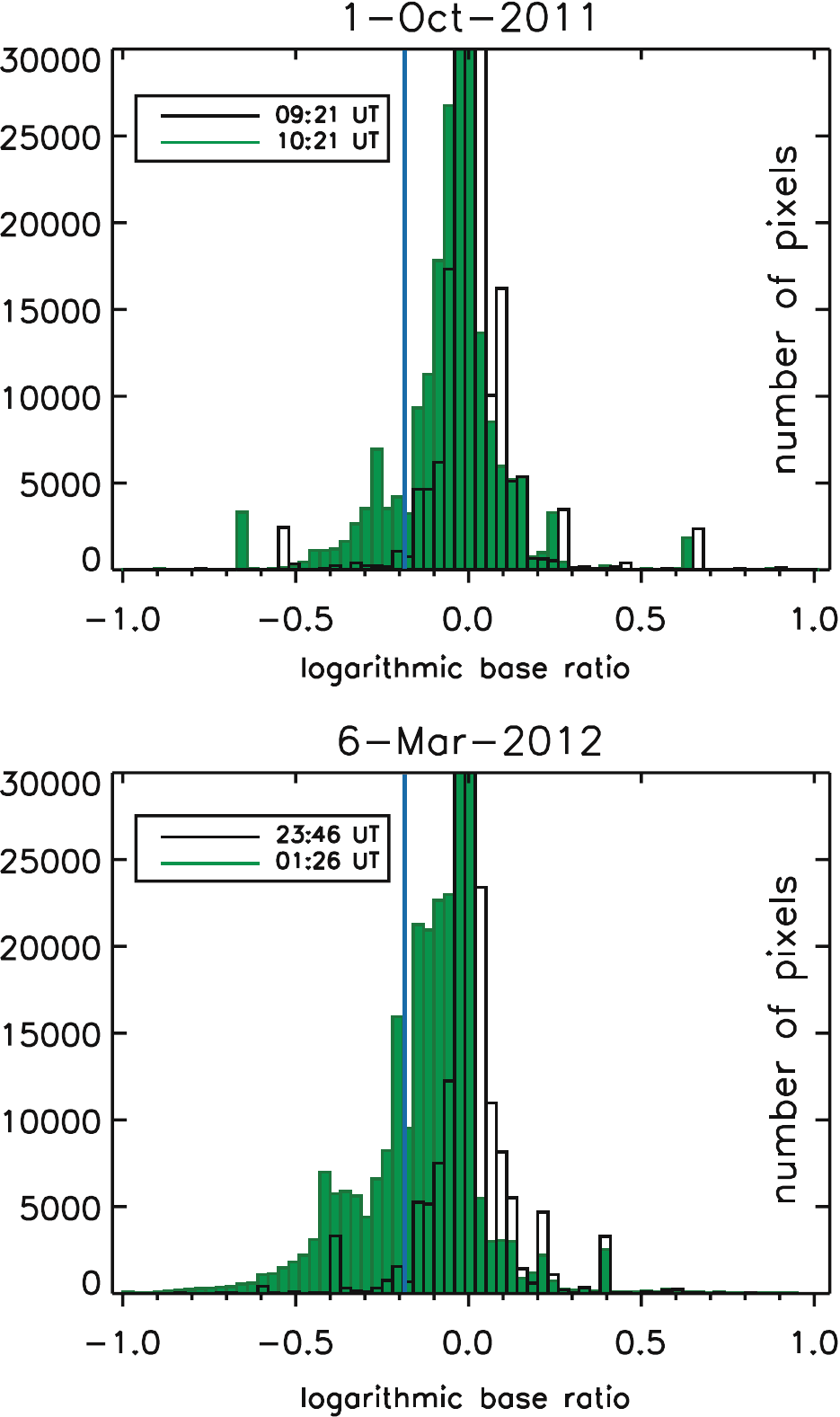}
\caption{Intensity distributions of logarithmic base-ratio images for dimming events that occurred on 2011 October 1 (top) and 2012 March 6 (bottom). The black histogram is calculated from an image during a pre-event time step (i.e. non-dimming state), and the green histogram is from an image during the time of maximal extent of the dimming. Note the shift toward negative intensity values for the distribution during the dimming state (green). The value used for first level thresholding is shown by the blue vertical line.}
\label{fig:histogram}
\end{figure}

\begin{figure*}[t]
\centering
\includegraphics[width=0.85\paperwidth]{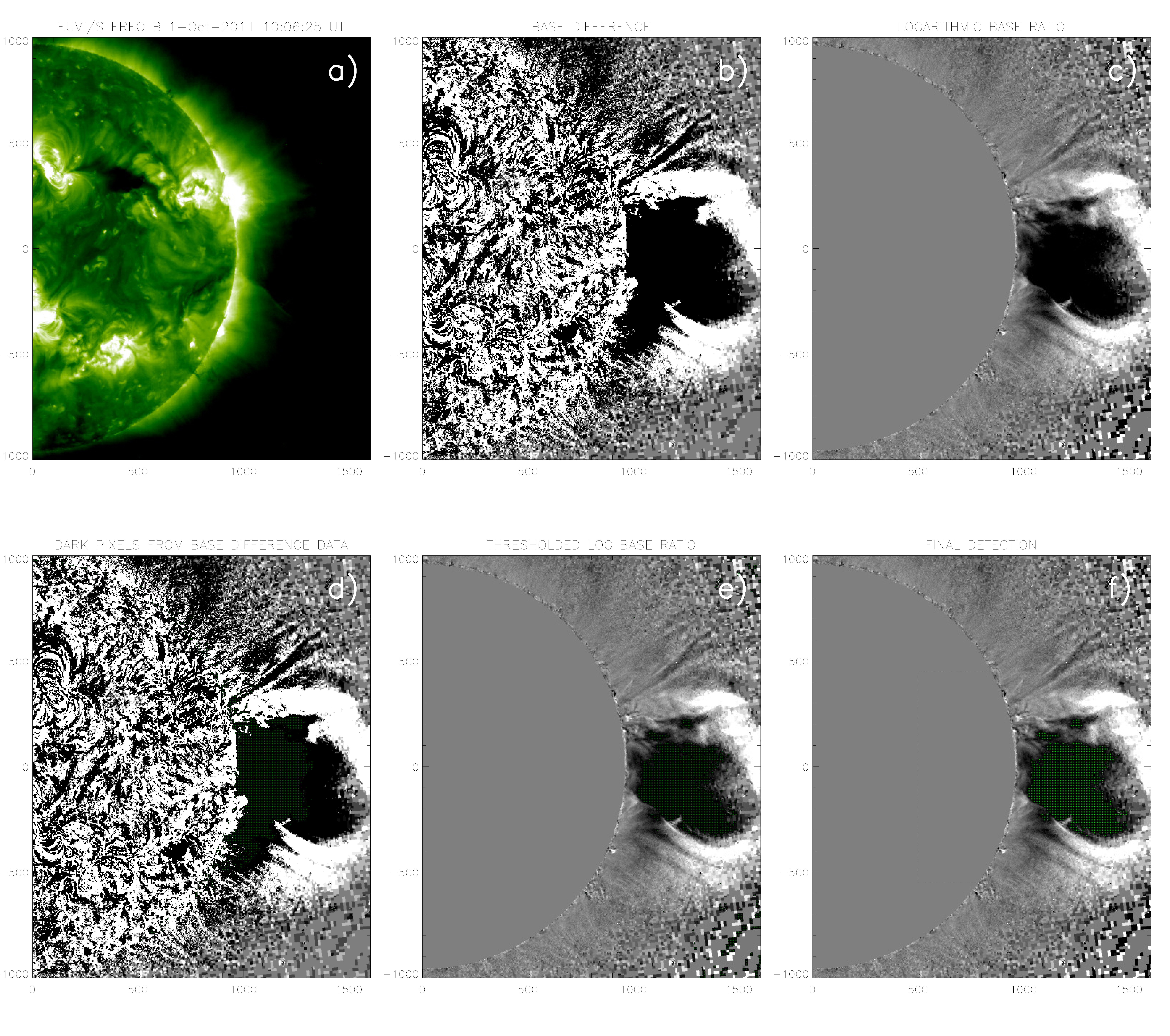}
\caption[width=0.85\paperwidth]{Dimming detection for the event on 2011 October 1. The top panels present original EUVI/STEREO 195 {\AA} filtegrams (a) together with base-difference (b) and logarithmic base-ratio (c) data. The bottom panels highlight pixels identified during the different steps of the detection: the darkest pixels extracted from base difference data (d), logarithmic base ratio data thresholded by level -0.19 (e), the final dimming region segmented for this time step in logarithmic base ratio data (f). The coordinates are given in arcsec from Sun center. The white rectangle on the last panel indicates the region shown in Fig.~\ref{cumulative_1}. }

\label{fig:detection example1}
\end{figure*}

In the next step, we select the thresholding level for logarithmic base ratio data: all pixels with a logarithmic base ratio intensity decrease below a value of -0.19 (which corresponds to a relative decrease by approximately 35\% in linear space) are identified as dimming pixels (cf. Fig.~\ref{fig:detection example1}e). Our threshold was found empirically from the histograms, by qualitatively checking different images during different development states of the dimming for several events \citep[see also][Fig.~2]{dissauer2018detection}. The thresholding level appeared to be the same as found in \cite{dissauer2018detection} for SDO/AIA data, as it performed well to separate the dimming pixel distribution from random noise within the majority of histograms investigated for the STEREO/EUVI data. In Figure~\ref{fig:histogram}, we show for two sample events, the intensity distribution of the logarithmic base ratio images for two different time steps, one before and one during the dimming. The state before the dimming formation is presented as black histogram and shows small variance in values; the dimming state at the time close to its maximal extent is presented as green histogram and reveals a systematic shift to negative intensity values, indicating the dimming pixels. The chosen thresholding level of -0.19 (blue vertical line) provides a basic, \textit{first level} separation of the non-dimming fluctuations in the image and the dimming pixels.

In order to reduce noise and the number of misidentiﬁed pixels, morphological operators (with a kernel of $3\times3$ pixels) are used to smooth the extracted regions and remove small-scale features. It also helps in situations, where gaps need to be filled in order to let seed pixels grow to more distant dimming parts. Finally, only pixels of LBR data, that represent direct neighbours of the seed pixels in BD data create the dimming region (cf. Fig.~\ref{fig:detection example1}f), and are subsequently used to extract the properties (area, intensity) of the dimming regions.

In Figure~\ref{fig:detection example1}, we illustrate all the major steps of the dimming detection algorithm. The top panels show the original data (a), the BD image (b) and the LBR image (c). The masks of the extracted regions for each step are shown in the bottom panels: the darkest pixels derived from the BD image (d), which are used as the seed pixels in the thresholded LBR data (e) by applying the region-growing algorithm to form the final dimming region (f). As it can be noticed, the noise in the LBR data is mostly present in regions higher up in the corona and does not overlap with the noise of BD data in the low corona. Thus, the chosen approach combining the two types of data is effective for the dimming extraction above the solar limb.

\subsection{Characteristic Coronal Dimming Parameters}

The set of parameters that we derive to characterize coronal dimmings and their evolution is extracted by the same approach as described in \cite{dissauer2018statistics}. We derive instantaneous and cumulative dimming areas from LBR data, instantaneous and cumulative brightness from both, BD and LBR data, their derivatives and the duration of the impulsive phase of the dimming, i.e. the time range when most of the dimming region develops.

\begin{figure*}[h]
\plotone{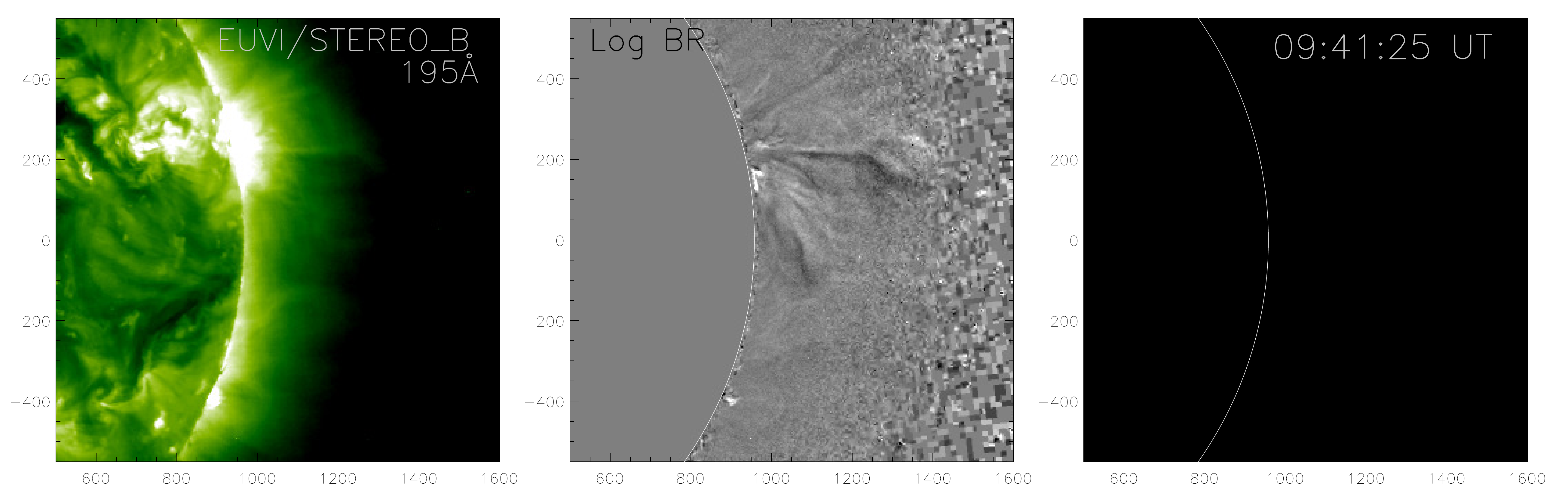}
\plotone{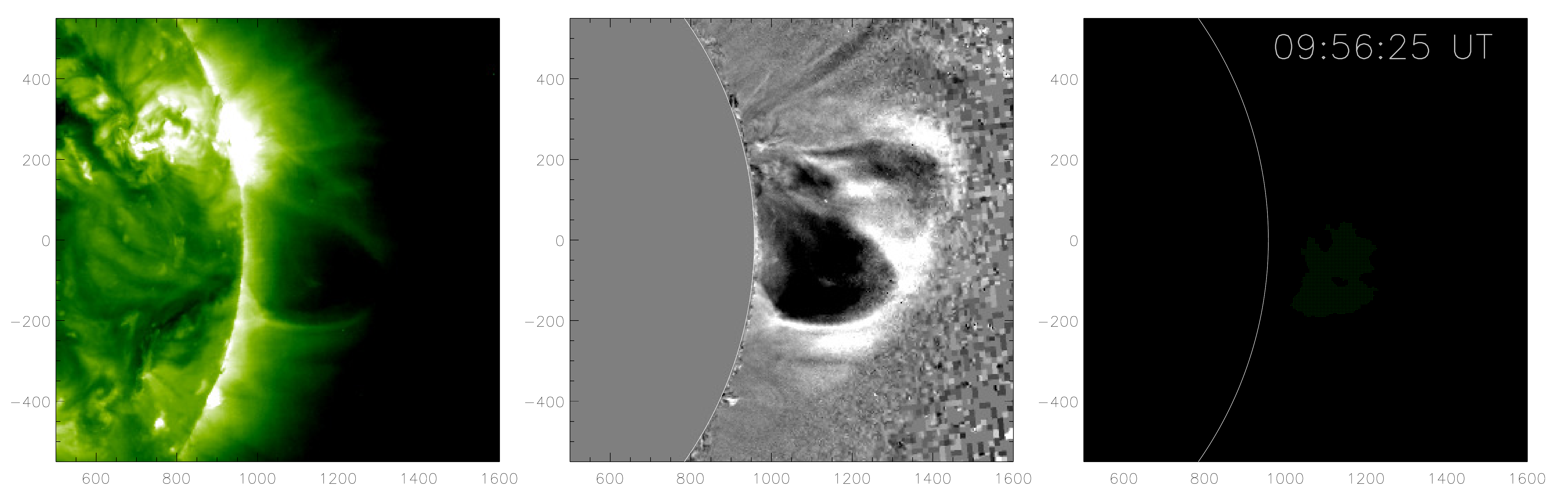}
\plotone{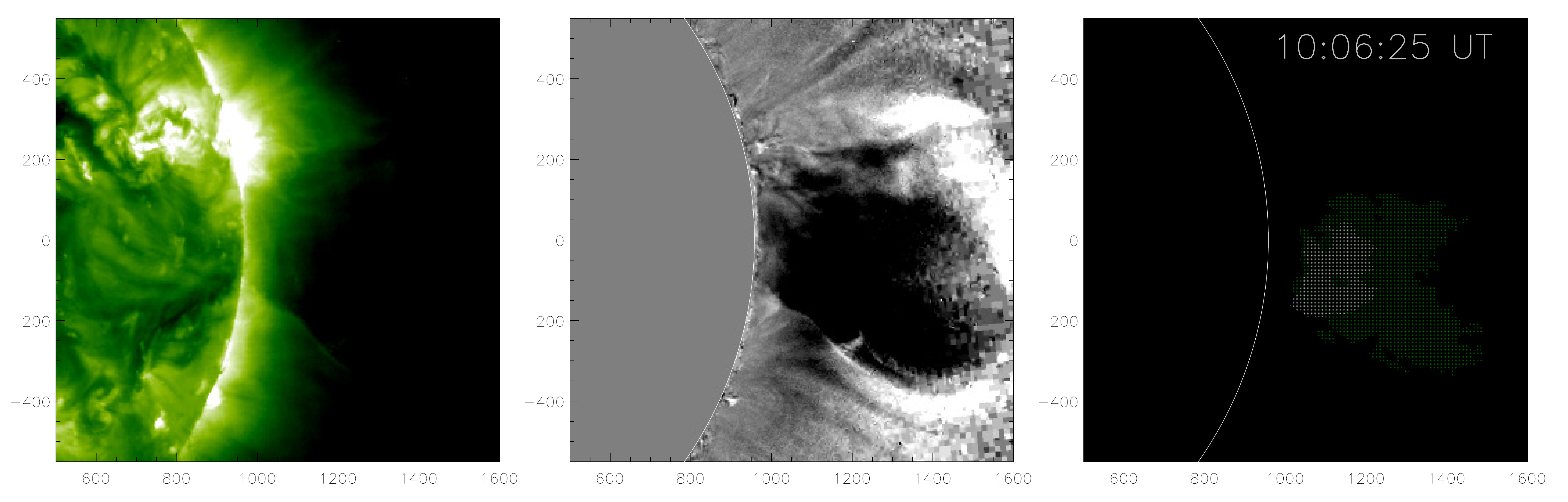}
\plotone{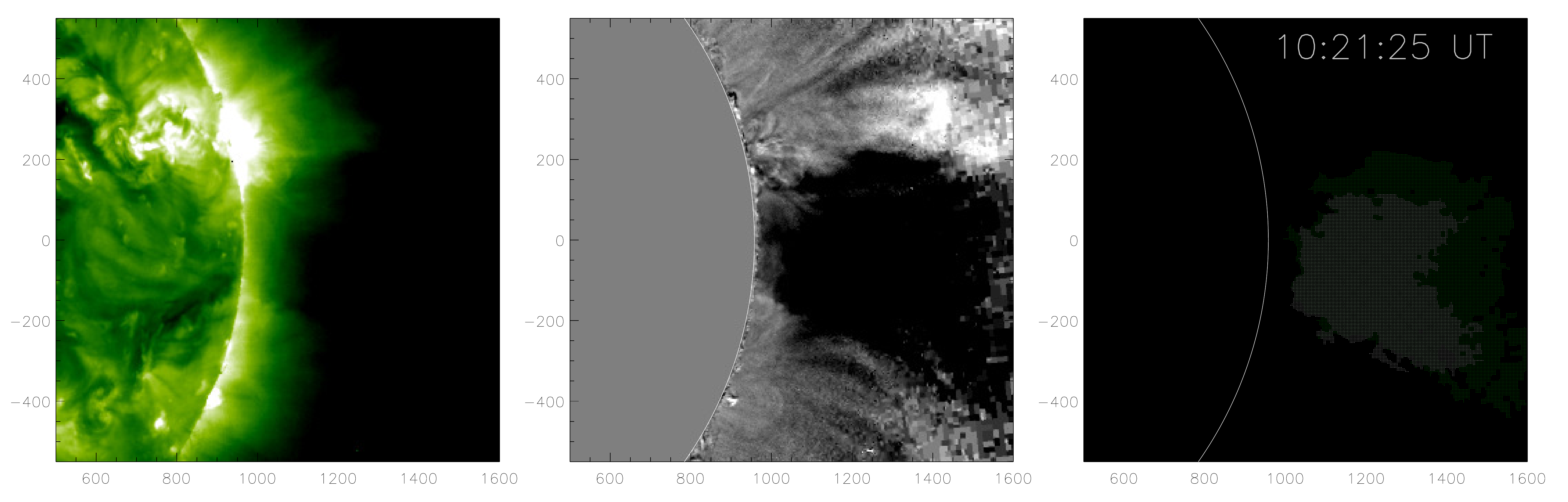}
\caption[width=1.0\textwidth]{Coronal dimming evolution for the 2011 October 1 event. From left to right: direct EUVI/STEREO-B filtergrams, logarithmic base-ratio images and cumulative dimming pixels masks. All the maps are zoomed to the region of the dimming location marked in Fig.\ref{fig:detection example1}. Right panels: the green pixels on top of the cumulative dimming pixel masks (grey regions) represent all newly detected pixels compared to the previously shown time step. The white contour in the middle and the right panels indicates the solar limb. The time of each detection is given in the right top corner of each panel. The values of the axes are given in arcsec.}
\label{cumulative_1}
\end{figure*}

\begin{figure*}
\plotone{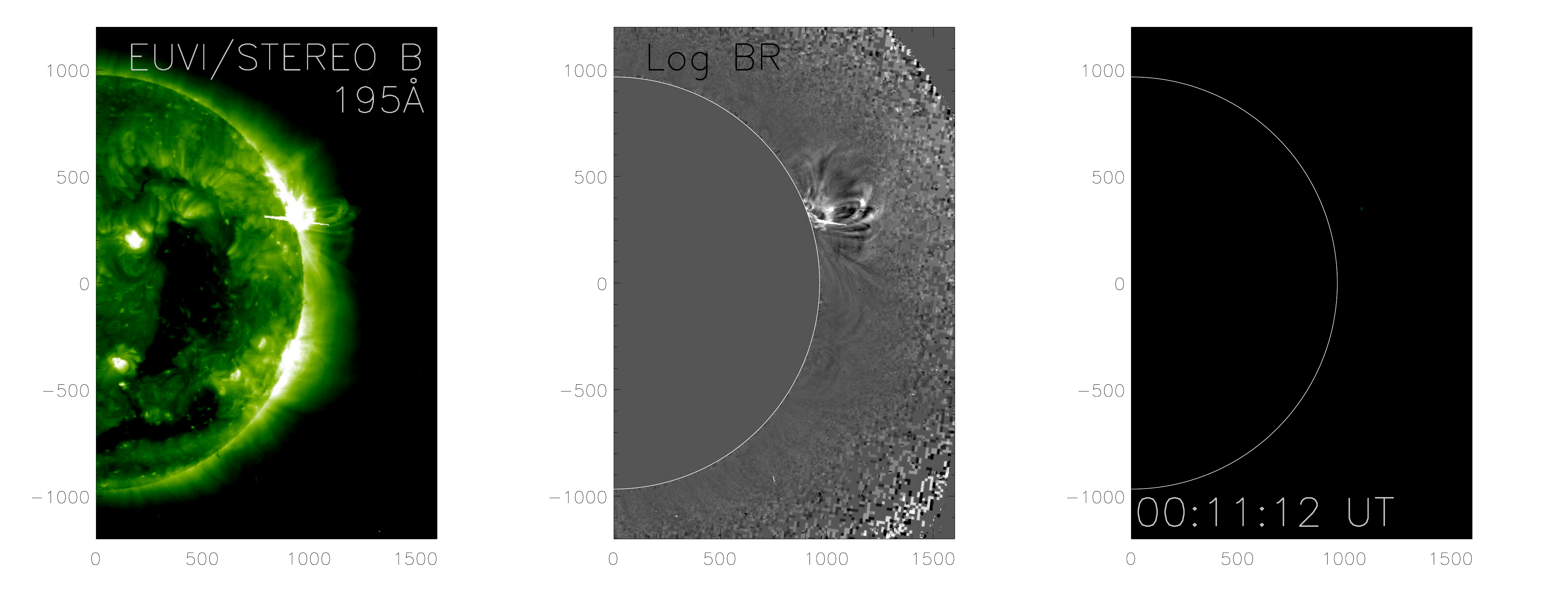}
\plotone{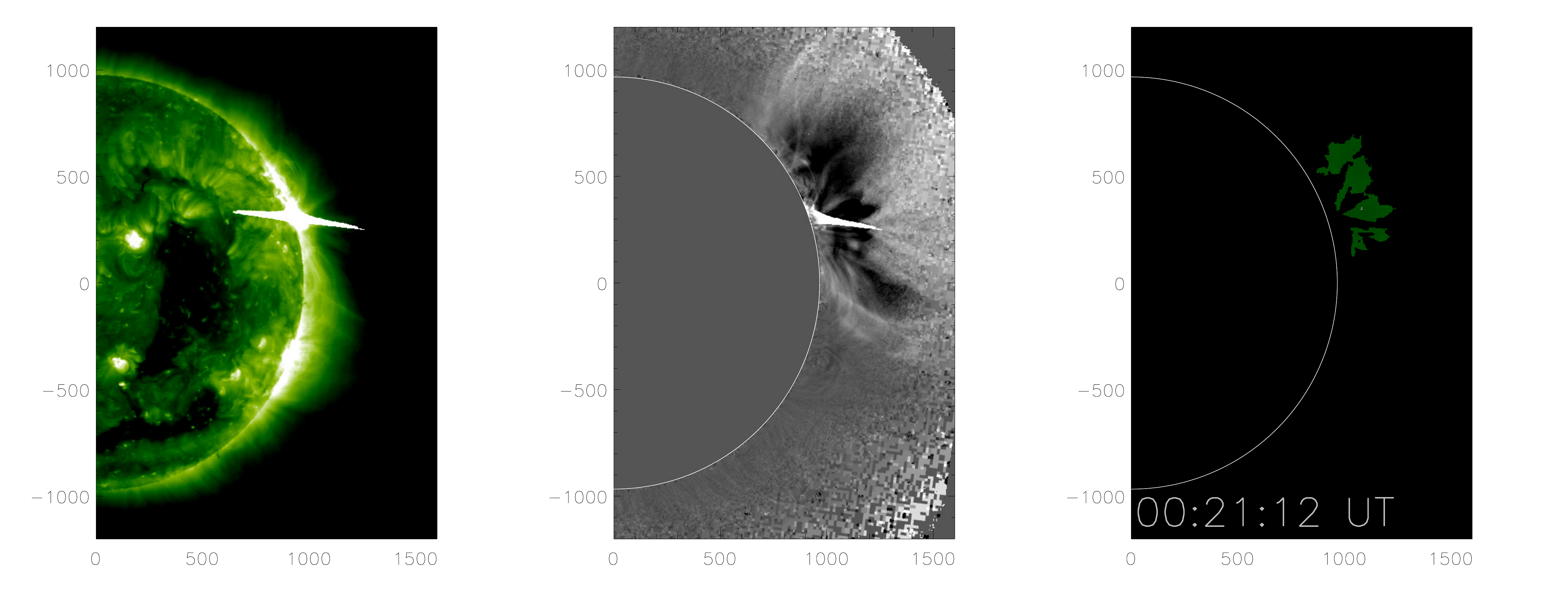}
\plotone{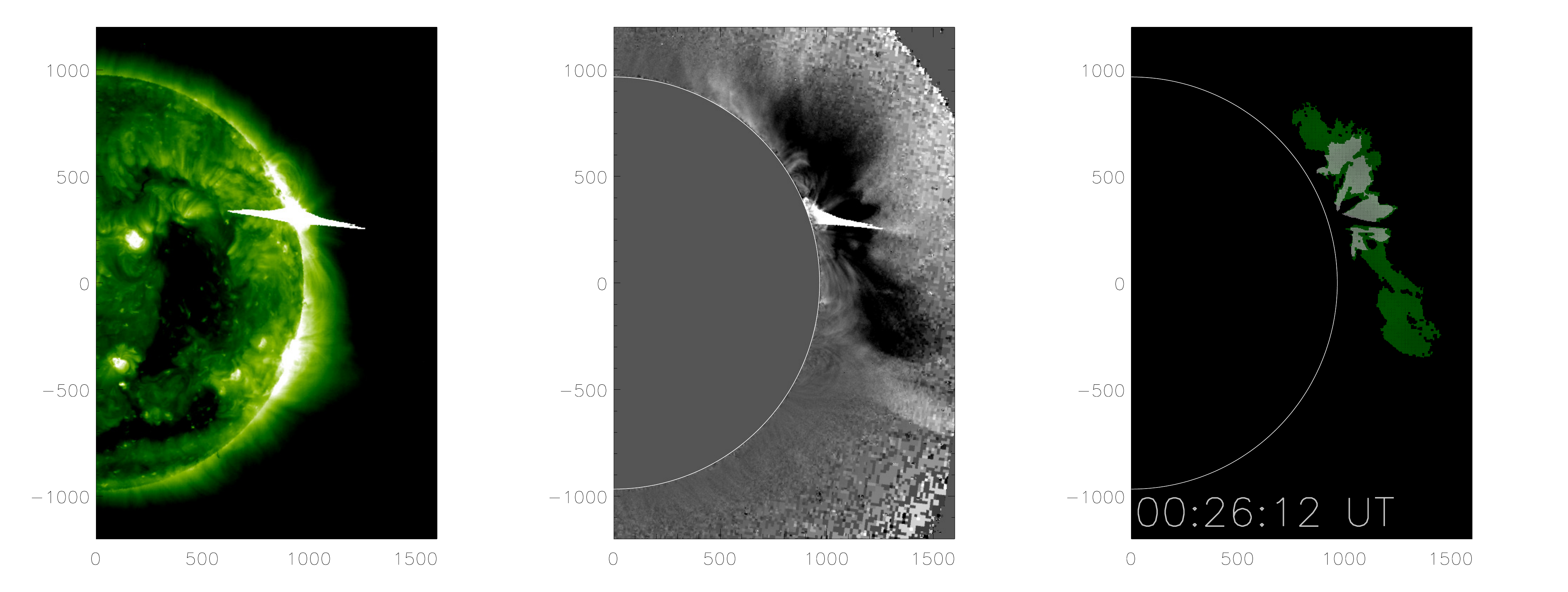}
\plotone{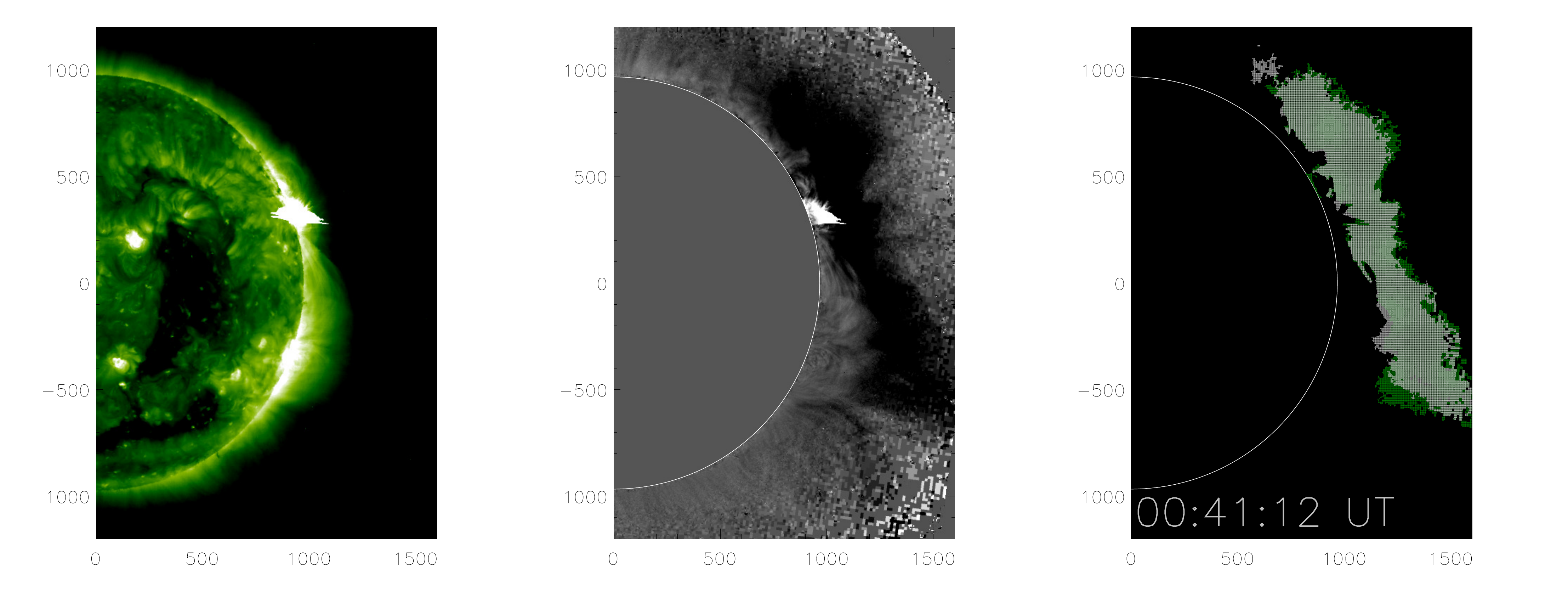}
\caption{Same as Fig.~\ref{cumulative_1} but for the 2012 March 6 event.}
\label{cumulative_2}
\end{figure*}

Instantaneous dimming masks are binary masks representing all dimming pixels detected at each time step. Cumulative dimming pixel masks are calculated by combining all the pixels, which were identiﬁed as dimming pixels up to a specific time. The choice of the approach for defining the dimming region influences the information we can extract from the evolution of the derived parameters. Cumulative masks allow us to study the full extent of the total dimming region over time. Figures~\ref{cumulative_1} and~\ref{cumulative_2} show the evolution of the cumulative dimming masks for the example events (\mbox{2011 October 1} and \mbox{2012 March 6}, respectively). The left and middle panels show snapshots of the original and logarithmic base-ratio STEREO/EUVI images. The corresponding dimming cumulative masks, extracted by our algorithm, are presented in the right panels. Green pixels show the newly detected dimming regions, while grey pixels represent all pixels which were extracted as part of the dimming region during previous time steps.

\begin{figure*}[h]
\centering
\includegraphics[width=0.75\textwidth]{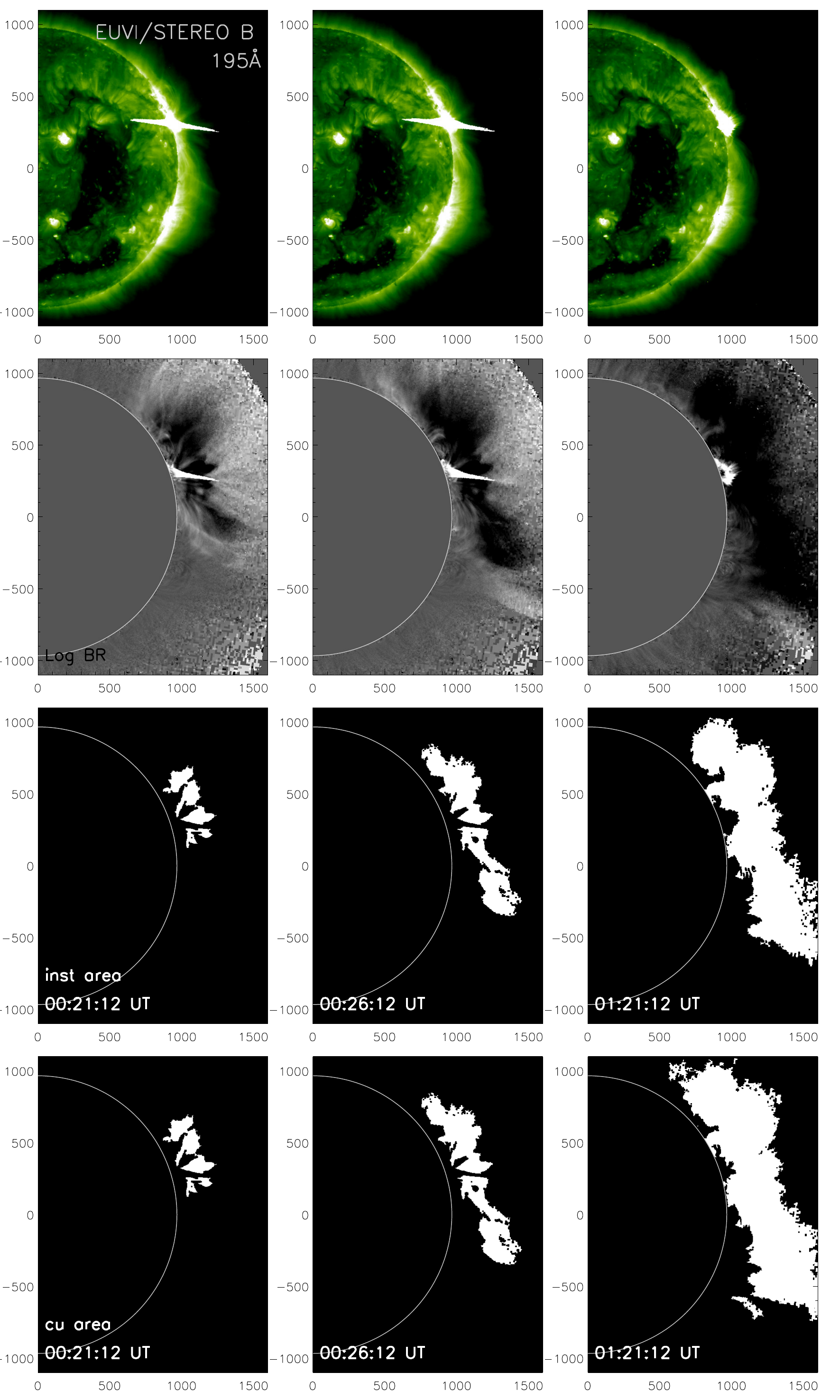}
\caption{Example of instantaneous and cumulative dimming masks for the 2012 March 6 event. Panels in the first row present original STEREO-B/EUVI filtegrams, while the second row shows the corresponding logarithmic base-ratio data. Panels in the third and fourth rows show instantaneous and cumulative dimming pixels masks, respectively. The snapshots shown correspond to the following phases during the dimming evolution: close to the start of the impulsive phase (left), the moment of the maximum area growth rate (middle), 20 minutes after the end of the impulsive phase (right). The white contour indicates the solar limb. The values of the axes are given in arcsec.}
\label{fig:time mask 2012}
\end{figure*}

\pagebreak

By using the extracted masks, the evolution of coronal dimming regions is studied by the instantaneous and cumulative areas: the instantaneous area $A_{in}(t_i)$ represents the number of pixels identified as dimming pixels at a certain time step $t_i$, the cumulative area $A_{cu}(t_i)$ is determined by combining the number of all dimming pixels detected up to time $t_i$. The corresponding area growth rate $dA/dt(t_i)$ characterizes how fast the dimming is developing; it is calculated as the corresponding time derivative of $A_{in}(t)$ or $A_{cu}(t)$ \citep[see also][]{dissauer2018detection}. The time evolution of these quantities allows us to study the dynamic evolution of the dimming region and to determine its impulsive phase.

The difference between instantaneous and cumulative masks, leading to a difference between the area parameters, can be seen in Figure~\ref{fig:time mask 2012}, which shows the development of the coronal dimming region for the event that occurred on \mbox{2012 March 6}. The start of the dimming impulsive phase (left), the time step of the maximum area growth rate (middle) and a time step 20 minutes after the end of the impulsive phase (right) are presented by the original STEREO/EUVI 195 {\AA} filtegrams (top), LBR data (second row), instantaneous (third row) and cumulative masks (fourth row). While the evolution of the event is rapid, the masks differ from each other mostly after the end of the impulsive phase of the dimming, when the cumulative area still contains the dimming parts extracted previously although it is already shrinking.

The total instantaneous brightness $I_{in}(t_i)$ of the dimming region at a certain time step is calculated as the sum of all dimming pixels intensities at the time $t_i$.
Instantaneous brightness values depend on both the area and the intensity of the pixels. In addition, to have a parameter which does not depend on the changing dimming area, we use one constant area $A$, representing the cumulative dimming mask at the end of the impulsive dimming evolution. The sum of pixel intensities from this area at each time step $t_i$ determines the cumulative brightness $I_{cu}(t_i)$. This parameter shows the intensity of all dimming pixels which were detected until the end time of the impulsive dimming evolution $t_{end}$, thus, it varies only because of the changes in the intensity of the dimming pixels and it does not depend on changes in the dimming area over time. The brightness change rate is given by the corresponding time derivative. Also, we define the mean intensity of the dimming region: $\overline{I}_{in}(t_i)$ by dividing the total instantaneous dimming brightness $I_{in}(t_i)$ by the instantaneous dimming area $A_{in}(t_i)$ at time $t_i$ and $\overline{I}_{cu}(t_i)$ by dividing the total cumulative brightness $I_{cu}(t_i)$ by the constant dimming area $A$.

Coronal dimmings start to develop co-temporal with the early evolution of CMEs and may remain for several hours after they were formed \citep[e.g.][]{dissauer2018statistics,2018ApJ...857...62V}. We study the impulsive phase of the coronal dimming evolution, defined by the area growth rate profile $dA/dt$. Following the definition in \mbox{\cite{dissauer2018detection, dissauer2018statistics}} we define the start of the impulsive phase $t_{start}$ as the local minimum that occurs closest in time before the highest peak of $dA/dt$. The end of the impulsive phase $t_{end}$ is defined as the time step when the value of the area growth rate falls below 15\% of its maximum:
\begin{equation}
\dot{A}(t) \leq 0.15 \cdot \dot{A}_{\rm max}.
\end{equation}
Thus the duration of the impulsive phase of the coronal dimming can be estimated as the difference between $t_{end}$ and $t_{start}$ (note, that the time cadence of STEREO/EUVI images is 5 minutes, which restricts the accuracy of the calculation of the duration parameter):
\begin{equation}
t_{\rm dur} = t_{end}-t_{start}.
\end{equation}
\label{eq}
\quad
To estimate the uncertainty of the extracted parameter values, caused by using a specific threshold for segmenting the dimming regions, we applied a $\pm5\%$ change to the logarithmic threshold level of -0.19 and then calculated the mean value and the standard deviation $\sigma$ for all the dimming parameters (see Figure~\ref{fig:evolution_of_parameters}). 

\section{Results}
We study a set of 43 coronal dimming events that are observed above the limb by the STEREO-A or STEREO-B EUVI instruments in the 195 {\AA} passband (cf. Table \ref{tab:table1}). These events have been observed during a period where the STEREO s/c were in quasi-quadrature with spacecraft located along the Sun-Earth line, and the same dimming events have been studied in observations against the solar disk by the SDO/AIA instrument in \mbox{\cite{dissauer2018statistics, dissauer2019statistics}}. For 37 events in our list, we have also the CME mass, and  for 27 events the CME maximum velocity derived from STEREO EUVI and COR data in \cite{dissauer2019statistics}. In Sect.~\ref{4.1} we present the temporal evolution of the dimming characteristics and their distribution. In Sect.~\ref{4.2} we relate our findings for the off-limb dimmings observed by EUVI to the results obtained for the corresponding on-disk dimming events observed by SDO/AIA from \cite{dissauer2018statistics}. In Sect.~\ref{4.3} we study the correlations of the decisive parameters describing off-limb coronal dimmings that we derived with the speed and mass of their associated CMEs. 

All the plots presented in the following sections show the mean values of the derived parameters, while the error bars represent the 1$\sigma$ standard deviation. The plots are presented in logarithmic space. The Pearson correlation coefficient $c$ as well as the linear regression fits shown are also derived in logarithmic space, i.e. $\log(y) = d + k \cdot \log(x)$. The values of the obtained correlation and fit parameters are annotated in each of the scatter plots shown. To obtain the mean and the standard deviation of $c$ we apply a bootstrap method \citep{wall2012practical}: we select random data pairs with replacements and repeat the procedures, calculating the Pearson coefficient 10000 times. In the same way as \citet{kazachenko2017database} and \citet{dissauer2018statistics} we classify the level of correlation as: $c=[0.2, 0.4]$ - weak, $c=[0.4, 0.6]$ - moderate, $c=[0.6, 0.8]$ - strong, $c=[0.8, 1.0]$ - very strong. 

\subsection{Evolution of the dimming parameters and their distribution} \label{4.1}

\begin{figure}
\epsscale{0.5}
\plotone{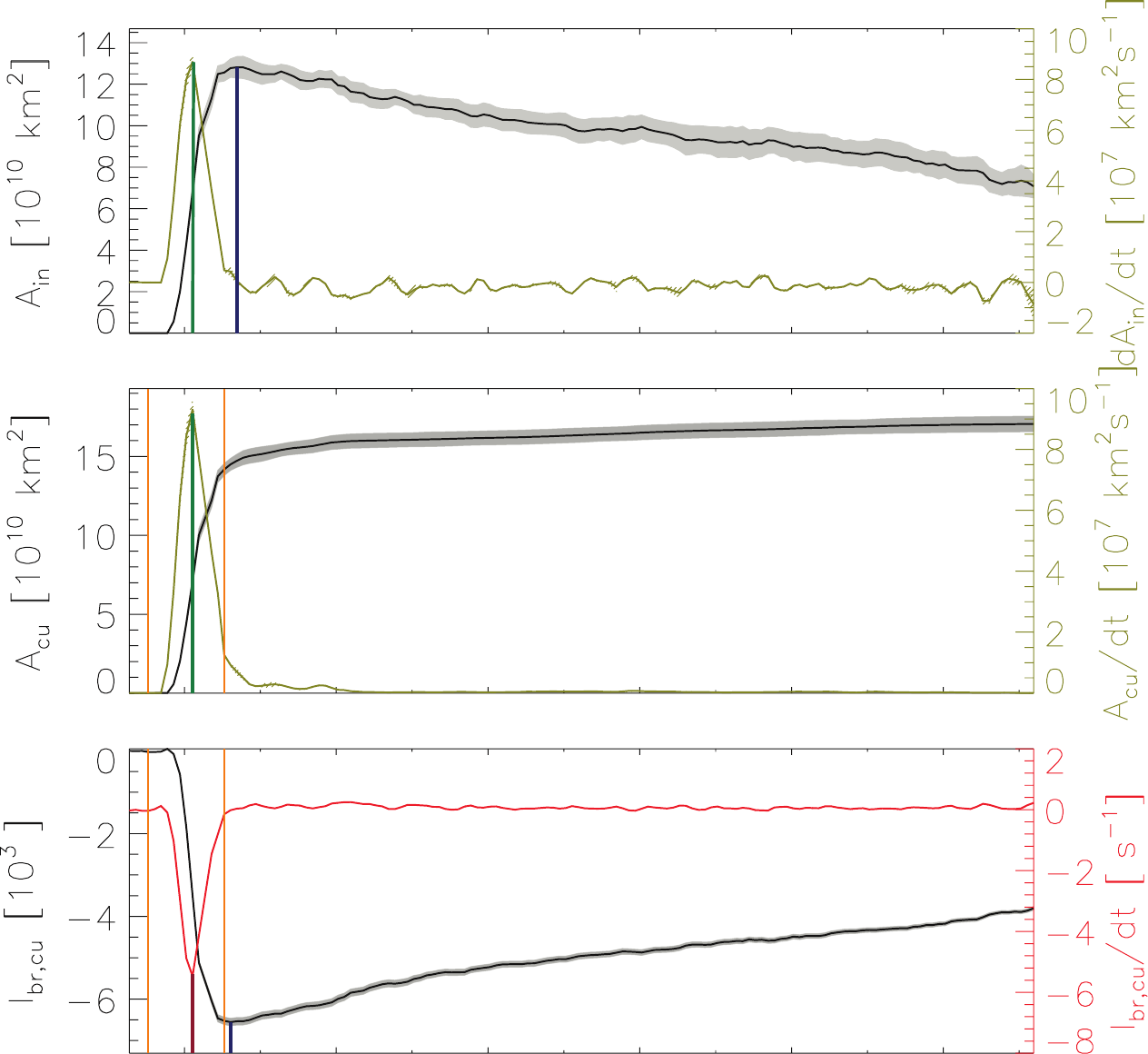}
\epsscale{0.52}
\plotone{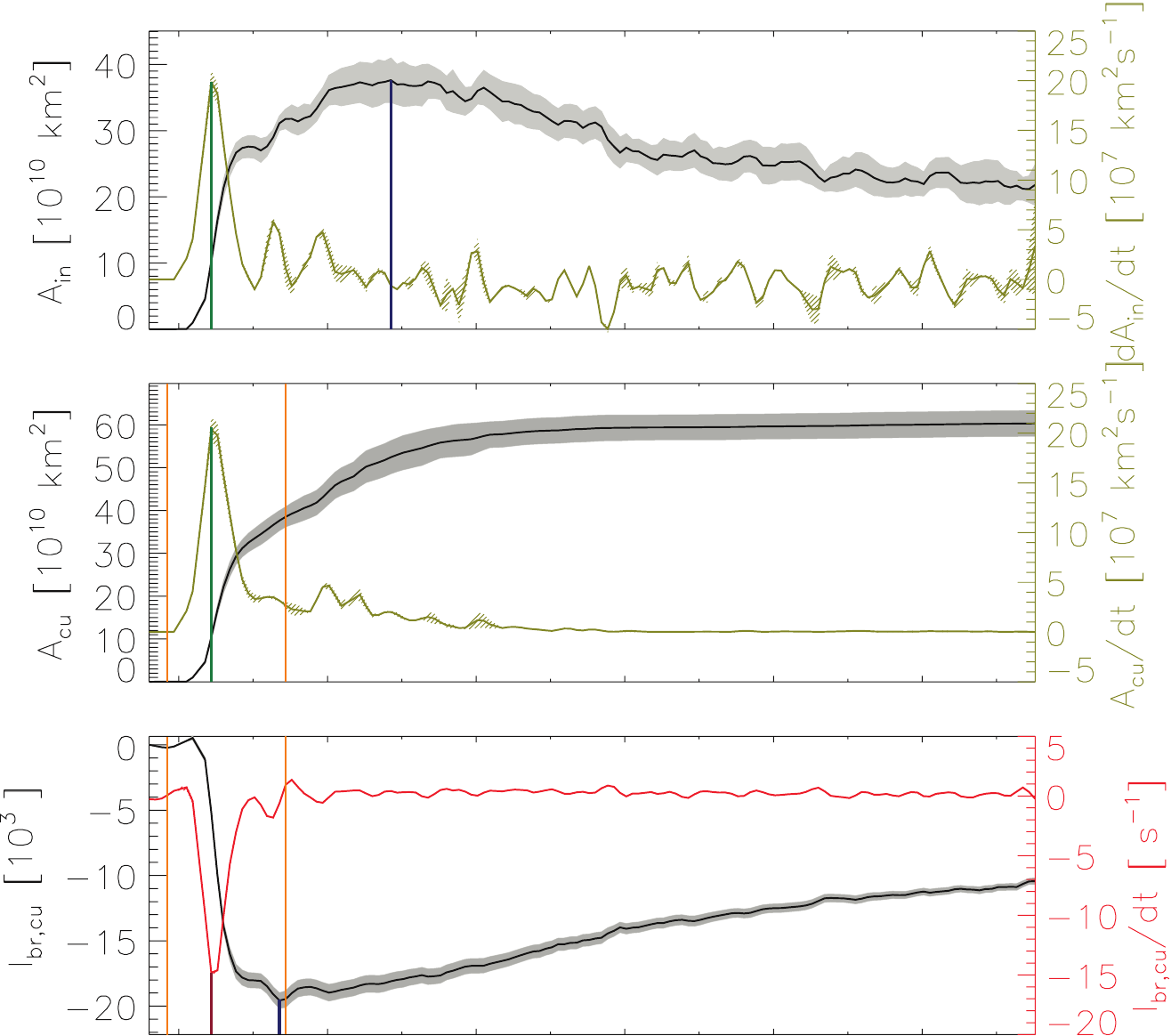}
\caption{Time evolution of the dimming parameters for the events that occurred on 2011 October 1 (left) and 2012 March 6 (right): (a) instantaneous area $A_{in}$ (black) and its growth rate (green), (b) cumulative area $A_{cu}$ (black) and its growth rate (green); (c) cumulative brightness from LBR maps I$_{br,cu}$ (black) and its change rates (red). The light gray bands show the uncertainty ranges for each parameter (1$\sigma$). Orange boundaries represent the start and the end time of the impulsive phase, green vertical lines show the maximum of the area growth rate, blue lines show the maximum of the instantaneous area or the minimum of the brightness, red line marks the minimum of the brightness change rate.}
\label{fig:evolution_of_parameters}
\end{figure}

\quad
Figure~\ref{fig:evolution_of_parameters} shows the time evolution of the coronal dimming parameters for two example events. Panels (a-b) show the instantaneous $A_{in}$ and the cumulative $A_{cu}$ area (black), respectively, together with the corresponding area growth rates (green). Panel (c) shows the total cumulative brightness $I_{br,cu}$ and its change rate (red). We note that the evolution of the total brightness is representative for the different brightness definitions, as they all are similar to each other, reaching their minimum at the same time (only 6 events from our data set reveal a significant difference in time of the minimum of instantaneous and cumulative brightness, one of them is the example event of 2012 March 6). The gray shadow bands represent the 1$\sigma$ range. Vertical lines indicate the start and the end time of the dimming impulsive phase (orange lines), the maximum of the area growth rate (green), the maximum of the instantaneous area and the minimum of the brightness (blue) and brightness change rate (red). In each event, the intensity decreases rapidly when the dimming area expands. The same tendency can be seen for the cumulative parameters, indicating that at this time the dimming regions become not only bigger, but also darker.

Studying the time evolution of the coronal dimming characteristics we extract parameters for the statistics and comparison with the CMEs quantities. We define the dimming size $A$ by the cumulative area at end of the impulsive phase of the dimming $t_{end}$. The maximum of the instantaneous area $A_{max}$ is defined as the largest size of the dimming region during the 6 hours after the start of the impulsive phase. The area growth rate parameters are obtained by the maximum of the $dA_{cu}(t)/dt$ and $dA_{in}(t)/dt$ profiles, respectively.

\begin{figure}[h]
\centering
\includegraphics[width=0.6\columnwidth]{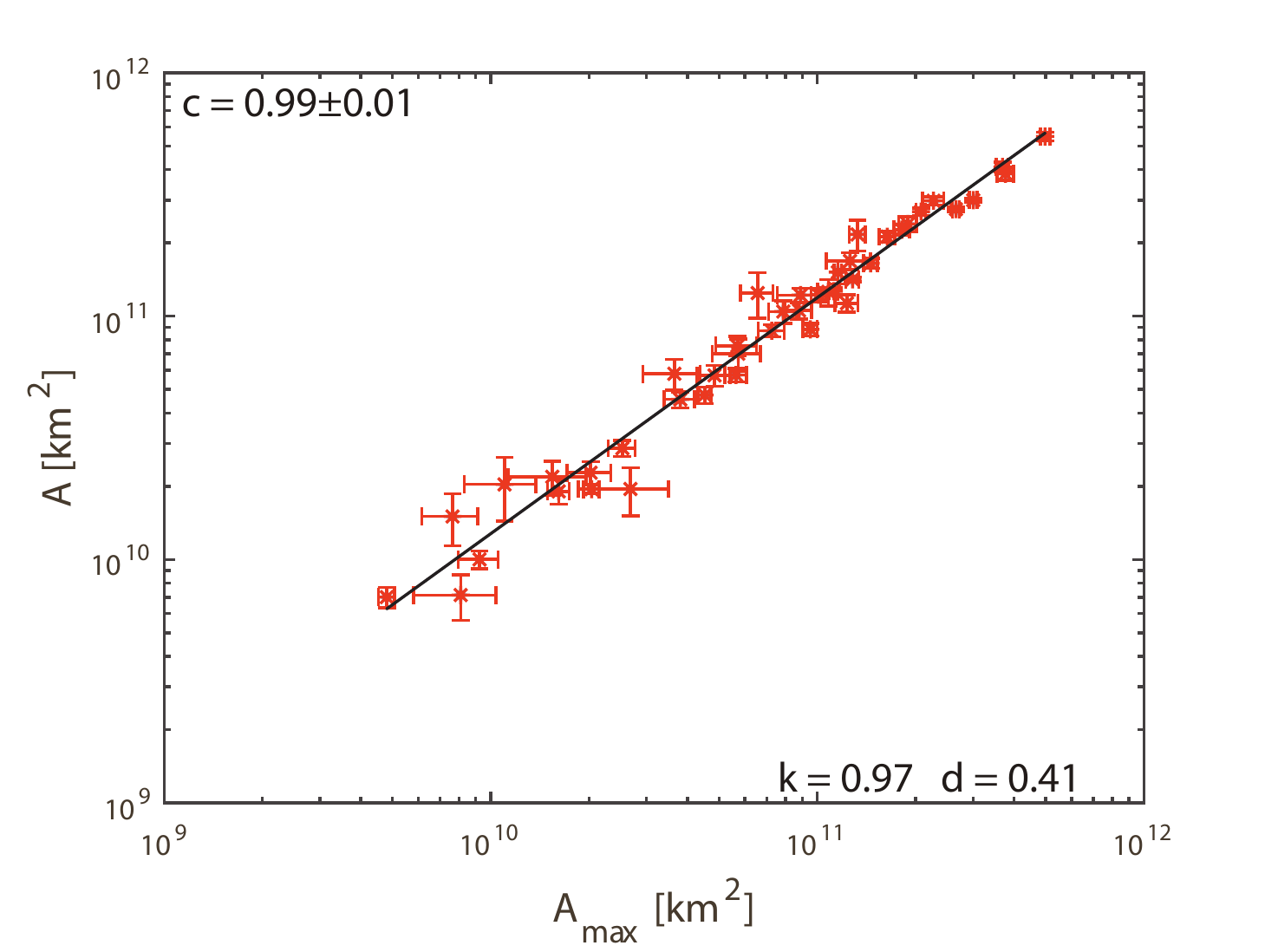}
\caption{The maximum of the instantaneous dimming area $A_{max}$ against the cumulative dimming area $A$ at the time $t_{end}$ of the impulsive dimming evolution (in logarithmic space). The black line represents the linear regression fit to all data points. The correlation coefficient is given in the top left corner. The parameters of the linear regression performed in log-log space are given in the bottom right corner.}
\label{fig:cu_max}
\end{figure}

Figure~\ref{fig:cu_max} compares the values of the parameters $A_{max}$ and $A$. The black line represents the linear regression fit to all data points. The coefficients of the fitting line are presented in the bottom-right corner ($k = 0.97$, $d = 0.41$). The corresponding correlation coefficient is given in the top-left corner ($c = 0.99\pm 0.01$). The high positive value of $c$ and the slope of the fitting regression line $k\sim1$ illustrates that the two different approaches for identifying the dimming area provide almost identical results. 

\begin{figure}[h]
\centering
\includegraphics[width=0.6\columnwidth]{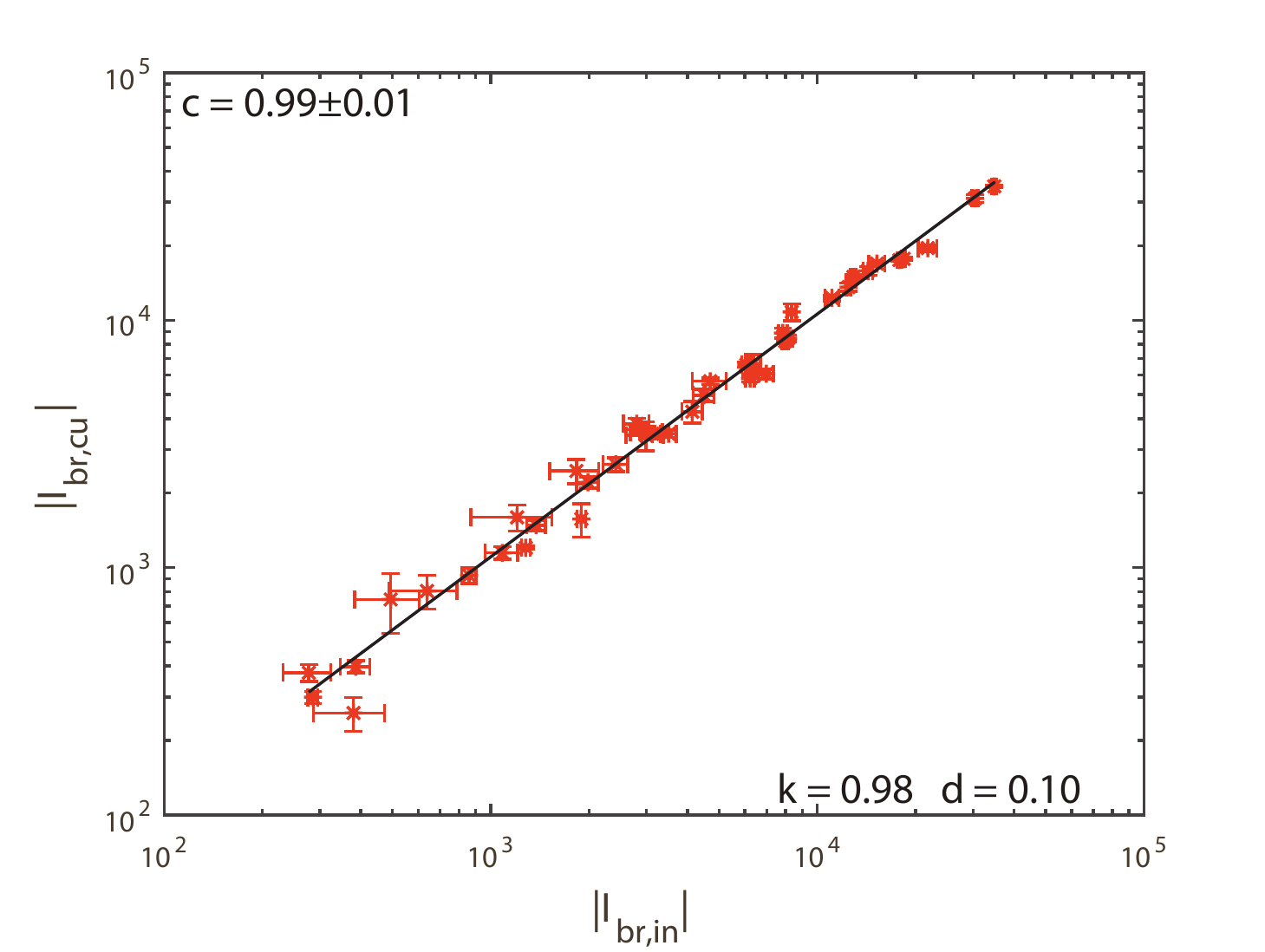}
\caption{Absolute values of the cumulative total brightness $|I_{br,cu}|$ versus absolute values of the instantaneous total brightness $|I_{br,in}|$ of the dimming regions, calculated from logarithmic base-ratio maps. The black line represents the linear regression fit to all data points. The coefficients of the fitting line are presented in the bottom right corner. The corresponding correlation coefficient is given in the top left corner.}
\label{fig:br_cu_in}
\end{figure}

We also extract the minimum of the total brightness for base-difference (instantaneous $I_{bd,in}$ and cumulative $I_{bd,cu}$) and logarithmic base-ratio (instantaneous $I_{br,in}$ and cumulative $I_{br,cu}$) data. Because $I(t)$ usually reaches the minimum after the dimming impulsive phase, we searched for their minimum in the time range of \mbox{6 hours} after $t_{start}$. The mean dimming brightness parameters $\overline{I}_{in}$ and $\overline{I}_{cu}$ are derived from the same time steps as the corresponding total brightness. Figure~\ref{fig:br_cu_in} shows that the instantaneous and cumulative brightness extracted from logarithmic base ratio maps (the same is also true for base-difference data) are almost identical \mbox{($c = 0.99$; $k = 0.98$, $d = 0.10$)}. This means that we can use one approach to describe the dimming brightness for the comparative analysis.

In Table~\ref{tab:table1}, we summarize all dimming parameters derived in our analysis. For each event we list the STEREO satellite, which was used for the analysis, and the derived dimming properties: the maximum of the instantaneous dimming area $A_{max}$, the cumulative area $A$, the maximal cumulative area growth rate $\dot{A}$, the total dimming brightness from logarithmic base-ratio data: cumulative $I_{\text{br,cu}}$ and instantaneous $I_{\text{br,in}}$, the total dimming brightness from base-difference data: cumulative $I_{\text{bd,cu}}$ and instantaneous $I_{\text{bd,in}}$, the mean instantaneous brightness $|\overline{I}_{br,in}|$, the brightness change rate $\dot{I}_{\text{br,in}}$ and the duration of the impulsive phase of the dimming $t_{dur}$. The CME quantities such as the mass $m_{CME}$ and the maximal speed $v_{max}$ are listed in the last columns of the table.

\begin{figure*}
\centering
\includegraphics[width=0.75\columnwidth]{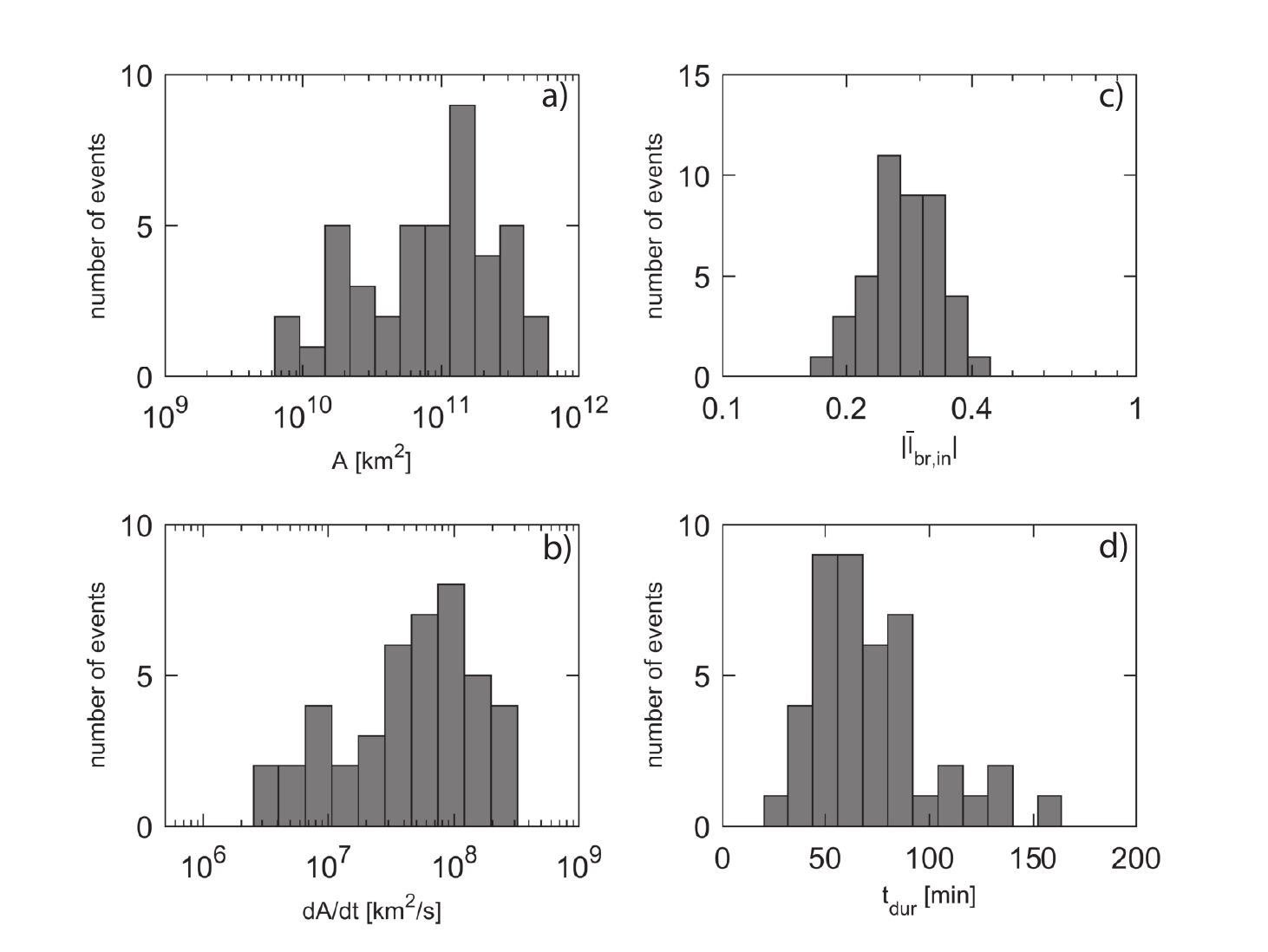}
\caption{Distributions of characteristic dimming parameters: a) maximum of the cumulative dimming area $A$, b) maximum of the cumulative area growth rate $dA/dt$, c) absolute values of the mean intensity from logarithmic base ratio data $|\bar{I}_{br,in}|$, d) duration of the dimming impulsive phase $t$. Panels (a-c) are presented in logarithmic space, panel (d) in linear space.}
\label{fig:hist_parameters}
\end{figure*}

Figure~\ref{fig:hist_parameters} shows the distributions of the main dimming parameters derived from the whole data set: the cumulative dimming area $A$ and its growth rate $dA/dt$, the absolute mean intensity from logarithmic base ratio data $\overline{I}_{br,in}$ and the duration of the impulsive phase $t_{dur}$ (note, that panels (a-c) are shown in the logarithmic scale). The dimming areas range from $7.0\times10^9$ km$^2$ to $5.5\times10^{11}$ km$^2$, with the mean value of $1.33\pm1.23\times10^{11}$ km$^2$. The area growth rate $dA/dt$ varies from $2.7\times10^6$ km$^2$ s$^{-1}$ to \mbox{$2.9\times10^8$ km$^2$ s$^{-1}$}, with the mean value of \mbox{$7.41\pm6.96\times10^{7}$ km$^2$ s$^{-1}$}. The mean brightness decrease of the total dimming regions varies in the interval [-0.41, -0.18], the mean value is $-0.28\pm0.05$ (note, that values of brightness are calculated from LBR data). The duration of the impulsive dimming phase varies from 25 to 155 minutes. On average, it lasts for $66\pm28$ minutes.

\pagebreak

\newcommand{\Msun}{{\rm M}_{\odot}}
\newcommand{\brems}{{bremsstrahlung}\xspace}
\newcommand{\Mvir}{{\rm M_{vir}}}
\newcommand{\Rvir}{{\rm R_{vir}}}
\newcommand{\Vvir}{{\rm V_{vir}}}
\newcommand{\Tvir}{{\rm T_{vir}}}
\newcommand{\LX}{{{\rm L_X}\,}}
\newcommand{\rhoc}{\rho_{\rm crit}}
\newcommand{\dd}{{\rm {d}}}
\newcommand{\DF}{{\small DF}\xspace}
\renewcommand{\tablename}{Table}
\begin{center}
\renewcommand{\thefootnote}{\fnsymbol{footnote}}
\renewcommand{\arraystretch}{1.0}
\setlength{\tabcolsep}{4.3pt}
\scriptsize

 \LTcapwidth=\textwidth
\fontsize{10}{16}\selectfont
\begin{longtable}{rcr|rrrrrrrrrr|rr} \hline

 \multicolumn{1}{c}{\text{$\#$}} & 
   \multicolumn{1}{c}{\text{Date}} &
    \multicolumn{1}{c}{\text{Sc}} &
   \multicolumn{1}{c}{\textbf{$A_{\text{max}}$}} &
   \multicolumn{1}{c}{\textbf{$A$}}  &
   \multicolumn{1}{c}{\textbf{$\dot{A}$}} &
   \multicolumn{1}{c}{\textbf{$I_{\text{br,cu}}$}} &
   \multicolumn{1}{c}{\textbf{$I_{\text{br,in}}$}}&
   \multicolumn{1}{c}{\textbf{$I_{\text{bd,cu}}$}} &
   \multicolumn{1}{c}{\textbf{$I_{\text{bd,in}}$}} &
   \multicolumn{1}{c}{\textbf{$\overline{I}_{\text{br,in}}$}} &
    \multicolumn{1}{c}{\textbf{$\dot{I}_{\text{br,in}}$}} &
    \multicolumn{1}{c}{\textbf{$t_{dur}$}}  &
   \multicolumn{1}{c}{\textbf{$m_{\text{CME}}$}}&
   \multicolumn{1}{c}{\textbf{$v_{\text{max}}$}} \\[0.8ex]

 \multicolumn{1}{c}{\text{  }} &
   \multicolumn{1}{c}{\text{  }} &
    \multicolumn{1}{c}{\text{  }} &
   \multicolumn{1}{c}{\textbf{[km$^{2}$]}} &
   \multicolumn{1}{c}{\textbf{[km$^{2}$]}} &
   \multicolumn{1}{c}{\textbf{[km$^{2}$s$^{-1}$]}} &
   \multicolumn{1}{c}{\textbf{  }} &
   \multicolumn{1}{c}{\textbf{}} &
   \multicolumn{1}{c}{\textbf{[DN]}} &
   \multicolumn{1}{c}{\textbf{[DN]}} &
   \multicolumn{1}{c}{\textbf{}} &
     \multicolumn{1}{c}{\textbf{[s$^{-1}$]}} &
   \multicolumn{1}{c}{\textbf{[min]}} & 
   \multicolumn{1}{c}{\textbf{[g]}} &
   \multicolumn{1}{c}{\textbf{[km s$^{-1}$]}} \\[0.8ex] 
   
   \multicolumn{1}{c}{\textbf{  }} &
   \multicolumn{1}{c}{\textbf{  }} &
    \multicolumn{1}{c}{\text{  }} &
  \multicolumn{1}{c}{\textbf{($\mathbf{10^{10}}$)}} &
   \multicolumn{1}{c}{\textbf{($\mathbf{10^{10}}$)}} &
   \multicolumn{1}{c}{\textbf{($\mathbf{10^{7}}$)}} &
   \multicolumn{1}{c}{\textbf{($\mathbf{10^{3}}$)}} &
   \multicolumn{1}{c}{\textbf{($\mathbf{10^{3}}$)}} &
   \multicolumn{1}{c}{\textbf{($\mathbf{10^{5}}$)}} &
   \multicolumn{1}{c}{\textbf{($\mathbf{10^{5}}$)}} &
    \multicolumn{1}{c}{\textbf{  }} &
        \multicolumn{1}{c}{\textbf{  }} &
    \multicolumn{1}{c}{\textbf{  }} &
       \multicolumn{1}{c}{\textbf{($\mathbf{10^{15}}$)}} &
        \multicolumn{1}{c}{\textbf{($\mathbf{10^{3}}$)}} \\[0.5ex]\hline \hline \\[-2ex]

\endfirsthead

1&20100716&B&0.76&1.50&0.40&-0.38&-0.28&-0.57&-1.20&-0.18&-0.11&120&-&0.37\\
2&20100801&A&22.68&29.81&11.73&-16.93&-15.22&-5.83&-6.00&-0.33&-6.89&100&4.05&1.26\\
3&20100807&A&8.79&10.56&6.30&-4.97&-4.50&-2.08&-1.84&-0.25&-2.84&55&6.69&0.96\\
4&20110213&A &7.86&10.45&3.77&-4.26&-4.14&-4.20&-3.27&-0.26&-1.48&105&2.50&1.02\\
5&20110214&A&0.48&0.70&0.27&-0.30&-0.29&-1.20&-1.51&-0.30&-0.07&65&0.47&-\\
6&20110215&A &5.69&7.56&2.98&-3.52&-3.01&-3.93&-3.25&-0.26&-1.56&80&6.27&1.33\\
7&20110307&B&1.54&2.19&0.80&-0.80&-0.64&-0.78&-0.40&-0.21&-0.30&90&4.34&1.10\\
8&20110621&A&16.37&21.22&8.27&-13.61&-12.46&-14.45&-13.72&-0.38&-7.39&77&6.97&1.00\\
9&20110711&A &9.52&8.83&5.71&-5.81&-6.22&-1.99&-3.56&-0.33&-3.19&50&1.48&0.53\\
10&20110802&A&11.24&12.67&9.66&-8.23&-7.97&-4.15&-4.66&-0.36&-5.95&40&6.98&-\\
11&20110803&A&18.14&22.68&6.62&-12.30&-11.08&-4.13&-4.80&-0.30&-3.11&105&7.69&1.61\\
*12&20110804&A&13.27&21.67&12.71&-10.80&-8.35&-3.12&-2.23&-0.34&-6.16&55&6.77&-\\
13&20110906&A &26.56&27.52&20.19&-17.70&-18.37&-7.00&-8.64&-0.34&-13.26&40&4.95&0.93\\
14&20110906&A&36.88&41.07&24.92&-31.07&-30.38&-20.46&-20.11&-0.41&-21.43&70&10.85&1.15\\
*15&20110907&A&20.76&26.91&16.00&-14.95&-12.89&-12.21&-11.39&-0.31&-9.66&55&5.35&-\\
16&20110908&A&0.81&0.71&0.45&-0.26&-0.38&-0.46&-0.72&-0.23&-0.17&55&0.22&0.37\\
17&20110927&B&0.92&1.00&0.61&-0.40&-0.39&-1.96&-1.73&-0.21&-0.23&65&-&-\\
18&20110930&A&2.02&2.28&0.88&-1.15&-1.09&-2.59&-2.41&-0.27&-0.37&95&-&-\\
19&20111001&B&12.81&14.17&9.32&-6.55&-6.24&-3.36&-3.18&-0.24&-4.22&60&-&0.57\\
20&20111002&A&3.65&5.80&2.63&-2.46&-1.83&-3.05&-2.87&-0.25&-1.35&65&2.43&0.84\\
21&20111002&A&1.61&1.91&1.06&-0.93&-0.86&-1.35&-1.28&-0.26&-0.55&70&1.73&0.63\\
22&20111124&A&10.09&12.24&2.99&-6.61&-6.07&-13.87&-14.26&-0.30&-1.69&155&3.02&-\\
23&20111222&A&3.81&4.55&2.79&-2.21&-1.99&-1.74&-1.70&-0.26&-1.34&55&2.37&-\\
24&20111225&A&2.68&1.95&1.13&-1.60&-1.20&-2.62&-1.42&-0.22&-0.39&40&-&-\\
25&20111225&A&5.73&7.01&7.03&-3.43&-2.98&-3.99&-3.81&-0.26&-3.36&40&5.90&-\\
26&20111226&A &14.56&16.50&8.86&-8.46&-8.14&-3.46&-3.43&-0.28&-4.05&60&4.60&1.02\\
*27&20120119&B&30.02&29.98&16.14&-17.50&-17.83&-5.64&-6.54&-0.30&-8.69&66&13.55&-\\
28&20120123&A &12.59&16.87&12.26&-8.88&-7.84&-6.27&-5.38&-0.31&-6.70&50&12.45&1.99\\
29&20120306&B&37.64&38.49&20.36&-19.57&-21.78&-11.71&-16.72&-0.29&-11.38&90&18.35&3.69\\
30&20120309&B&7.25&8.71&6.85&-3.47&-3.13&-2.80&-2.49&-0.21&-2.92&30&7.02&1.25\\
31&20120310&A&18.93&23.93&12.13&-15.84&-14.29&-9.96&-10.85&-0.38&-9.22&70&10.83&1.65\\
32&20120314&A &5.64&5.73&3.61&-3.48&-3.51&-3.62&-3.62&-0.31&-2.16&50&3.41&-\\
33&20120317&A&2.04&1.95&1.82&-1.20&-1.28&-1.15&-1.43&-0.31&-0.94&30&0.20&-\\
34&20120405&A&11.56&15.17&9.15&-8.53&-7.97&-2.38&-2.10&-0.34&-4.42&50&5.95&-\\
35&20120511&A &2.53&2.87&0.75&-1.48&-1.38&-1.57&-1.53&-0.27&-0.43&130&3.37&1.16\\
36&20120603&B&12.32&11.30&10.46&-6.08&-6.98&-1.57&-2.39&-0.29&-5.26&35&3.68&-\\
37&20120606&A&4.84&5.71&4.58&-2.61&-2.42&-0.96&-0.80&-0.25&-2.09&40&3.13&0.77\\
38&20120614&A &10.80&12.59&7.23&-6.65&-6.36&-3.76&-4.02&-0.29&-3.71&65&7.76&1.44\\
39&20120712&B &49.80&54.77&29.76&-34.76&-34.81&-7.31&-7.06&-0.35&-18.16&55&17.80&1.27\\
*40&20120804&B&6.56&12.45&3.58&-3.81&-2.80&-4.70&-3.38&-0.21&-1.24&25&11.05&-\\
41&20120815&A&1.10&2.04&1.38&-0.74&-0.49&-0.30&-0.15&-0.22&-0.43&50&1.03&0.60\\
42&20120925&B&4.52&4.75&3.35&-1.57&-1.89&-0.73&-1.57&-0.21&-1.04&35&-&0.47\\
43&20120927&A&8.88&12.21&7.47&-5.67&-4.70&-3.33&-2.71&-0.26&-4.00&80&9.37&1.50\\*[0.3ex]\hline \hline
\\[-2ex]

\caption[Dimming characteristics]{For each event we list the STEREO satellite used for the analysis and the derived dimming characteristics: the dimming instantaneous area $A_{max}$, the cumulative area $A$, the maximal cumulative area growth rate $\dot{A}$, the total dimming brightness from logarithmic base ratio data: cumulative $I_{\text{br,cu}}$ and instantaneous $I_{\text{br,in}}$, the total dimming brightness from base difference data: cumulative $I_{\text{bd,cu}}$ and instantaneous $I_{\text{bd,in}}$, the mean instantaneous brightness $|\overline{I}_{br,in}|$, the brightness change rate $\dot{I}_{\text{br,in}}$ and the duration of the impulsive phase of the dimming $t_{dur}$. Also the CMEs quantities are given: the mass of the CME $m_{CME}$ and the maximal speed of the CME $v_{max}$. Events marked with * are not included in the on-disk dimming study by \cite{dissauer2018statistics,dissauer2019statistics}.}\label{tab:table1} 

\end{longtable}

\normalsize
\end{center}

\begin{figure*}[h]
\centering
\includegraphics[width=0.58\columnwidth]{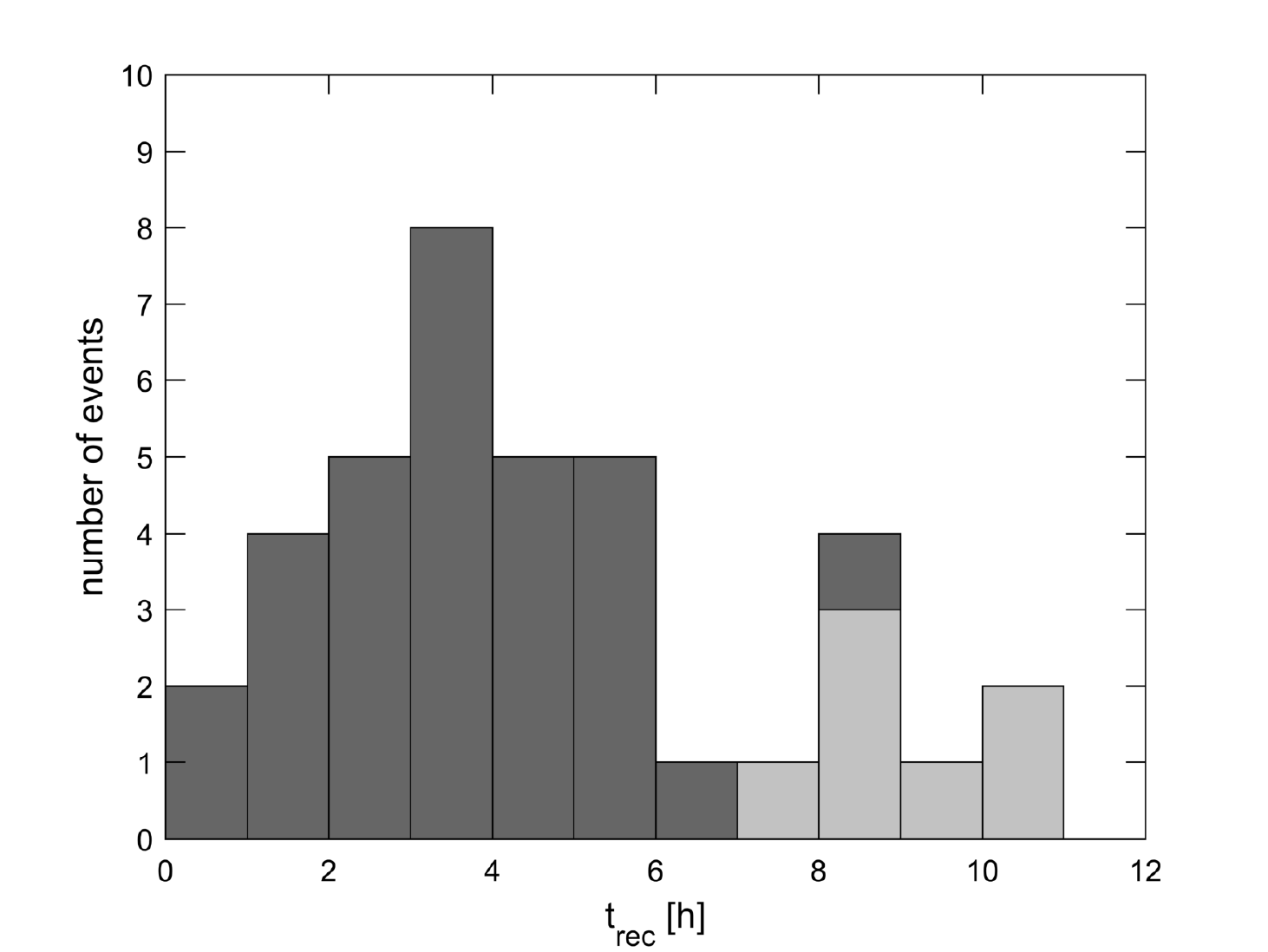}
\caption{Distribution of dimming recovery times derived from \textbf{38} events of our sample. Note that for 7 events (light coloured bars), the derived recovery time gives only a lower estimate. }
\label{fig:time_rec}
\end{figure*}

In addition to the characteristic dimming parameters listed above, we estimated the recovery time of the dimmings, which indicates how fast the corona restructures and refills after the erupting CME. The recovery time $t_{rec}$ is defined as the difference between the time of the maximal instantaneous area $A_{in}(t)$ ( = $A_{max}$) and the moment when the value of the instantaneous dimming area $A_{in}(t)$ falls below 50\% of its maximum ( = $0.5\cdot A_{max}$). For 5 events it was impossible to calculate the recovery time because of subsequent CME eruptions from the same active region during the dimming recovery. The distribution of the parameter $t_{rec}$ for the other 38 dimming events is plotted in Figure \ref{fig:time_rec}. The recovery time varies from 0.7 hrs to 10.9 hrs with a mean value of $4.6\pm2.8$ hrs. However, for 7 events of the sample, the derived recovery times are underestimated, as the drop to a value of $0.5\cdot A_{max}$ was not reached within the 12 hrs length of the time series that we studied for each event. We note that this includes the two example events. The histogram seems to indicate a bimodal distribution with one group  of fast recovery (smaller than $\sim$ 6 hrs), and another group with longer recovery times of $>$ 6 hrs. To provide a definite conclusion on that, a larger sample would be needed, and to follow the dimmings over a longer time range. The analysis of the dimming recovery time is important as it provides information on the post-eruptive phase of the CME and the coronal restructuring.

\subsection{Relation between off-limb and on-disk dimmings} \label{4.2}

\begin{figure}[H]
\centering
\includegraphics[width=0.6\columnwidth]{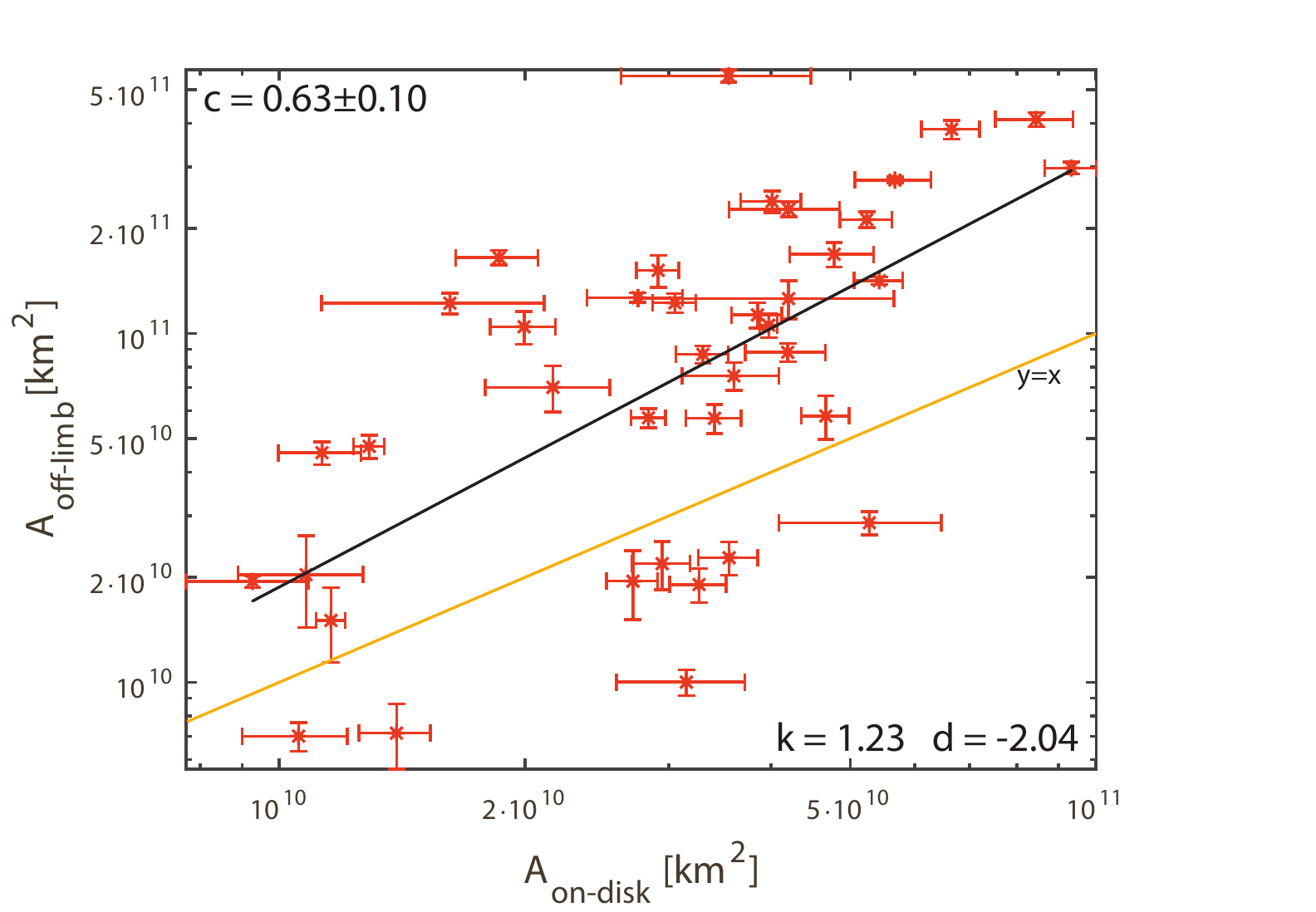}
\caption{Scatter plot of the dimming areas derived off-limb by STEREO/EUVI, $A_{\text{off-limb}}$, and on-disk by SDO/AIA, $A_{\text{on-disk}}$ in logarithmic space. The black line represents the linear regression fit to all data points. The orange line shows the one-to-one relation. The corresponding correlation coefficient is given in the top left corner. The parameters of the linear regression performed in log-log space are given in the bottom right corner.}
\label{fig:area-on-off}
\end{figure}

We compare the dimming parameters that we obtained from the STEREO/EUVI 195 {\AA} off-limb observations with the corresponding outcomes for the dimmings observed on-disk by SDO/AIA 211 {\AA} in \cite{dissauer2018statistics,dissauer2019statistics}. Figure~\ref{fig:area-on-off} shows the scatter plot between the dimming areas derived from on-disk and off-limb observations, which reveal a strong correlation ($c = 0.63 \pm 0.10$). For the events under study, off-limb dimmings tend to be larger than their corresponding on-disk counterpart: only 8 events reveal a smaller dimming area for the off-limb observations compared to the on-disk areas.

\begin{figure}[H]
\centering
\includegraphics[width=0.5\columnwidth]{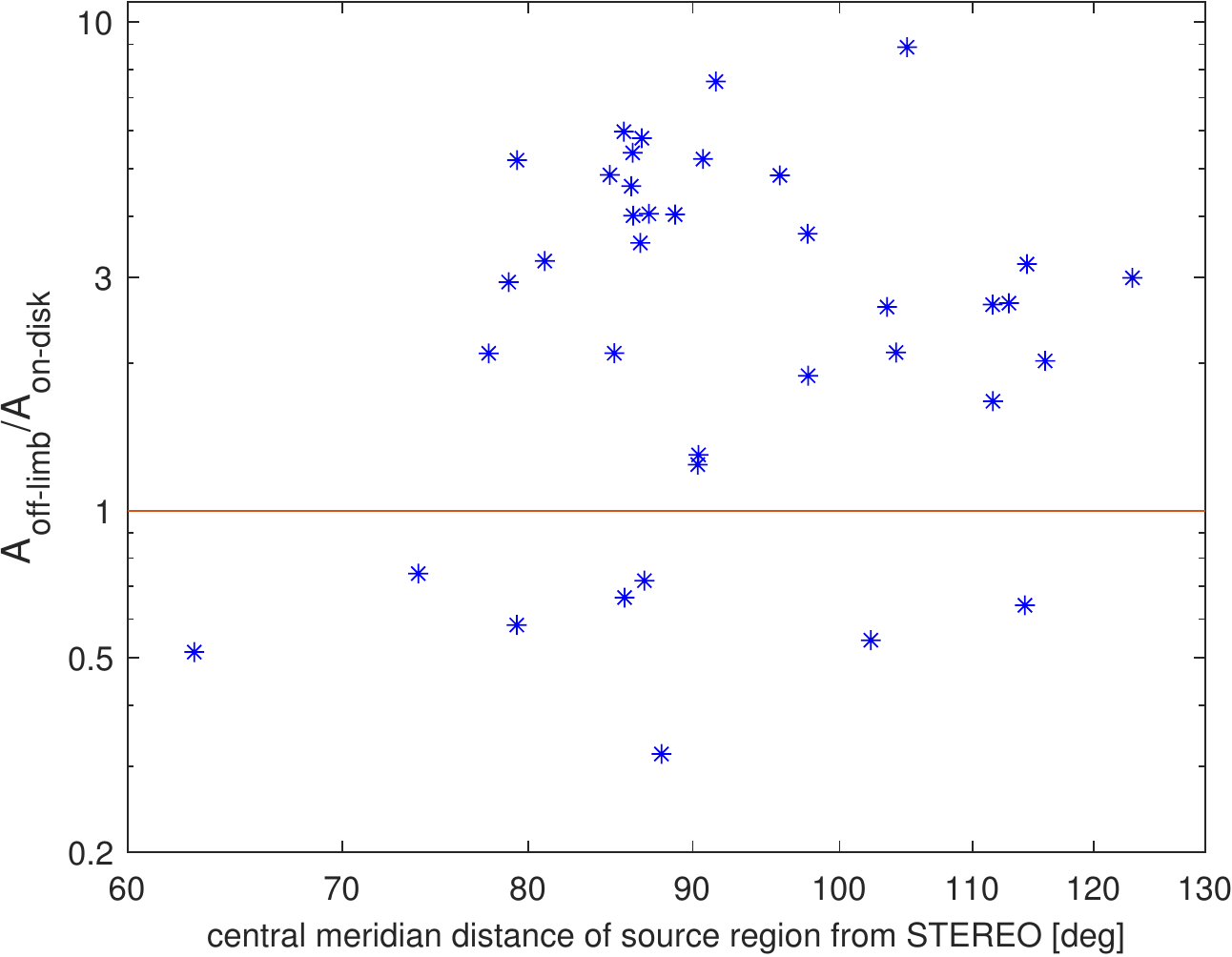}
\caption{Ratio of the extracted off-limb and on-disk dimming areas versus the central meridian distance of the source region from STEREO satellite. The orange line corresponds to one-to-one relation.}
\label{fig:position}
\end{figure}

In order to investigate, whether the position of the STEREO satellites significantly affects the derived dimming area via projection effects or by obscuring parts of the dimming region behind the limb, we studied the dimming areas as function of source region location with respect to the observing spacecraft. Figure \ref{fig:position} shows the ratio of the off-limb area $A_{off-limb}$ and on-disk area $A_{on-disk}$ against the central meridian distance of the CME source region for the STEREO spacecraft. The coordinates of the dimming sources are derived from the heliographic positions of the associated flares given in  \cite{dissauer2018statistics}. We can notice, that all the events were observed with a central meridian distance (CMD) in the range [60$^{\circ}$, 130$^{\circ}$], but there is no significant dependence of the area ratio between the off-limb and on-disk observations on the CMD. Also, there is no obvious change at a CMD of $>90^{\circ}$, i.e. for events where for STEREO the associated flare is located behind the limb, and thus part of the dimming regions may be obscured.

\begin{figure}[H]
\centering
\includegraphics[width=0.6\columnwidth]{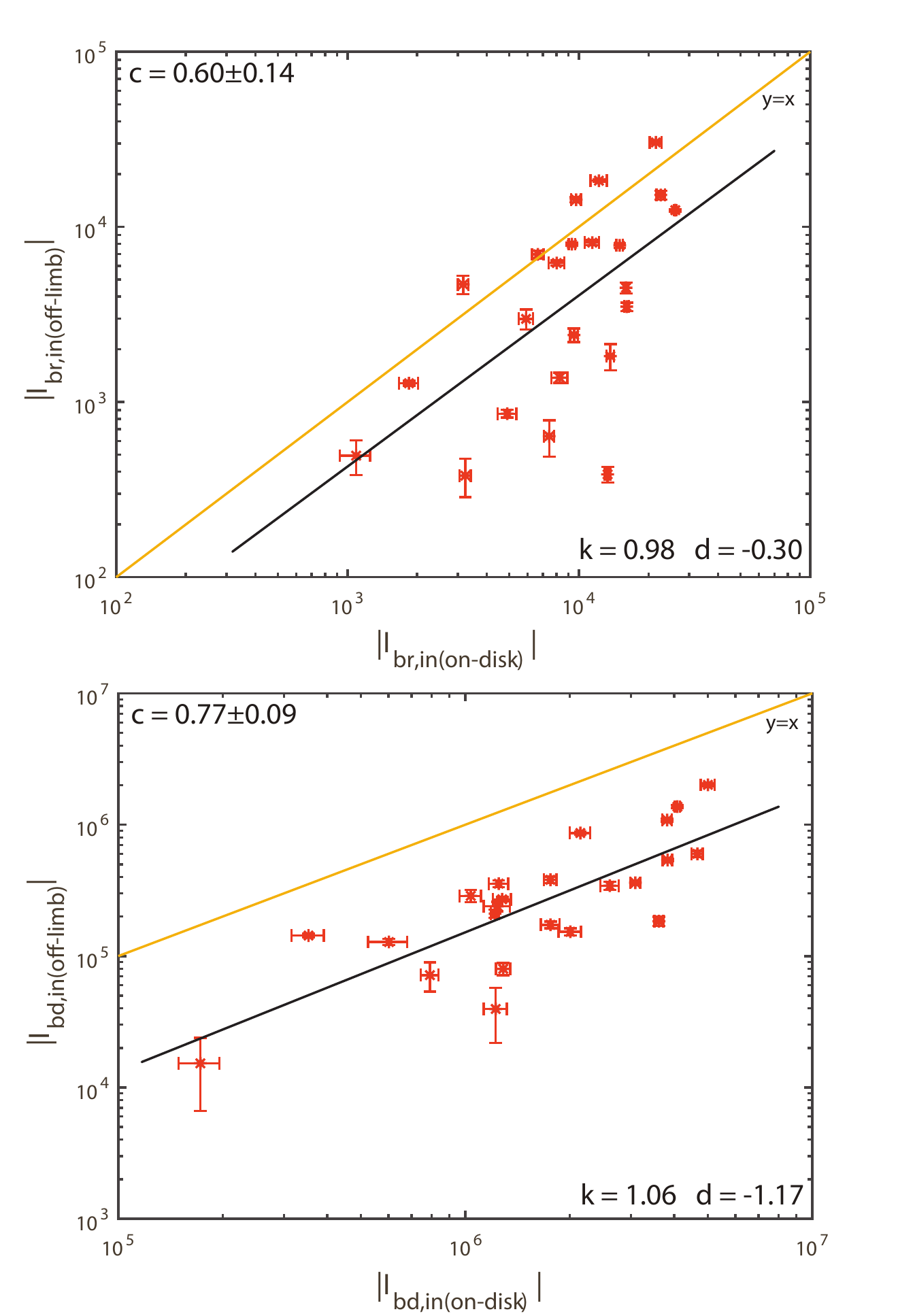}
\caption{Comparison of the total instantaneous brightness from logarithmic base-ratio data (top panel) and from base-difference data (bottom panel) for on-disk and off-limb observations, presented in absolute values. The black line represents the linear regression fit to all data points. The orange line shows the one-to-one relation. The corresponding correlation coefficients are given in the top left corner of each panel. The parameters of the linear regression performed in log-log space are given in the bottom right corner.
}
\label{total-on-off}
\end{figure}

Figure \ref{total-on-off} shows the scatter plot of total instantaneous brightness calculated from off-limb and on-disk observations by using logarithmic base-ratio (top panel) and base-difference (bottom panel). The correlation coefficients result in $c = 0.60 \pm 0.14$ and $c = 0.77 \pm 0.09$, respectively. The regression lines (in black) have a slope coefficient close to 1 ($k = 0.98$ and $k = 1.06$) and are therefore almost parallel to the 1:1 correspondence lines (orange), indicating that both parameters are linearly related. 

\subsection{Relation between off-limb dimming and CME parameters} \label{4.3}

\begin{figure}[H]
\centering
\includegraphics[width=0.6\columnwidth]{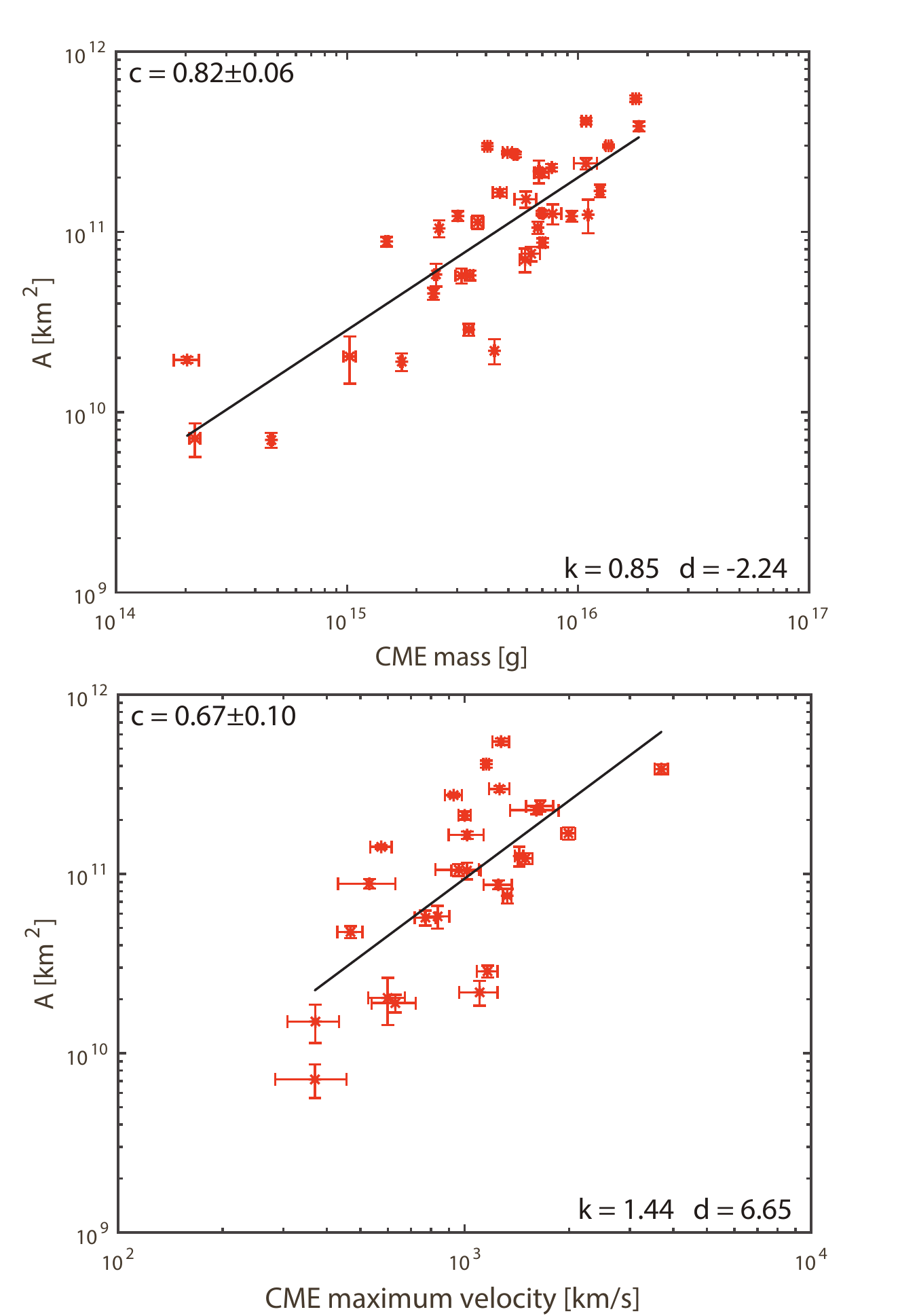}
\caption{Cumulative area of the coronal dimming $A$ against the CME mass $m_{CME}$ (top) and the CME maximal speed $v_{max}$ in logarithmic space. The black line represents the linear regression fit to all data points. The corresponding correlation coefficient is given in the top left corner. The parameters of the linear regression performed in log-log space are given in the bottom right corner.}
\label{fig:area_mass}
\end{figure}

Figure \ref{fig:area_mass} shows the cumulative off-limb dimming area $A$ against the mass (top panel) and maximum speed (bottom panel) of the CME. We obtain a very high correlation between the dimming area and CME mass: $c = 0.82\pm 0.06$ (in logarithmic space), i.e. the larger the area of the dimming, the more mass the associated CME contains. This correlation provides strong support of the physical interpretation of the appearance of the dimming as a density depletion due to the evacuation of plasma. These results are in agreement with the findings for the on-disk coronal dimmings by \cite{dissauer2019statistics}, where $c$ = $0.69\pm 0.10$.
We also find a high correlation between the area of the dimming $A$ and the speed of the CME ($c = 0.67\pm 0.10$ in logarithmic space).

\begin{figure}[H]
\centering
\includegraphics[width=0.6\columnwidth]{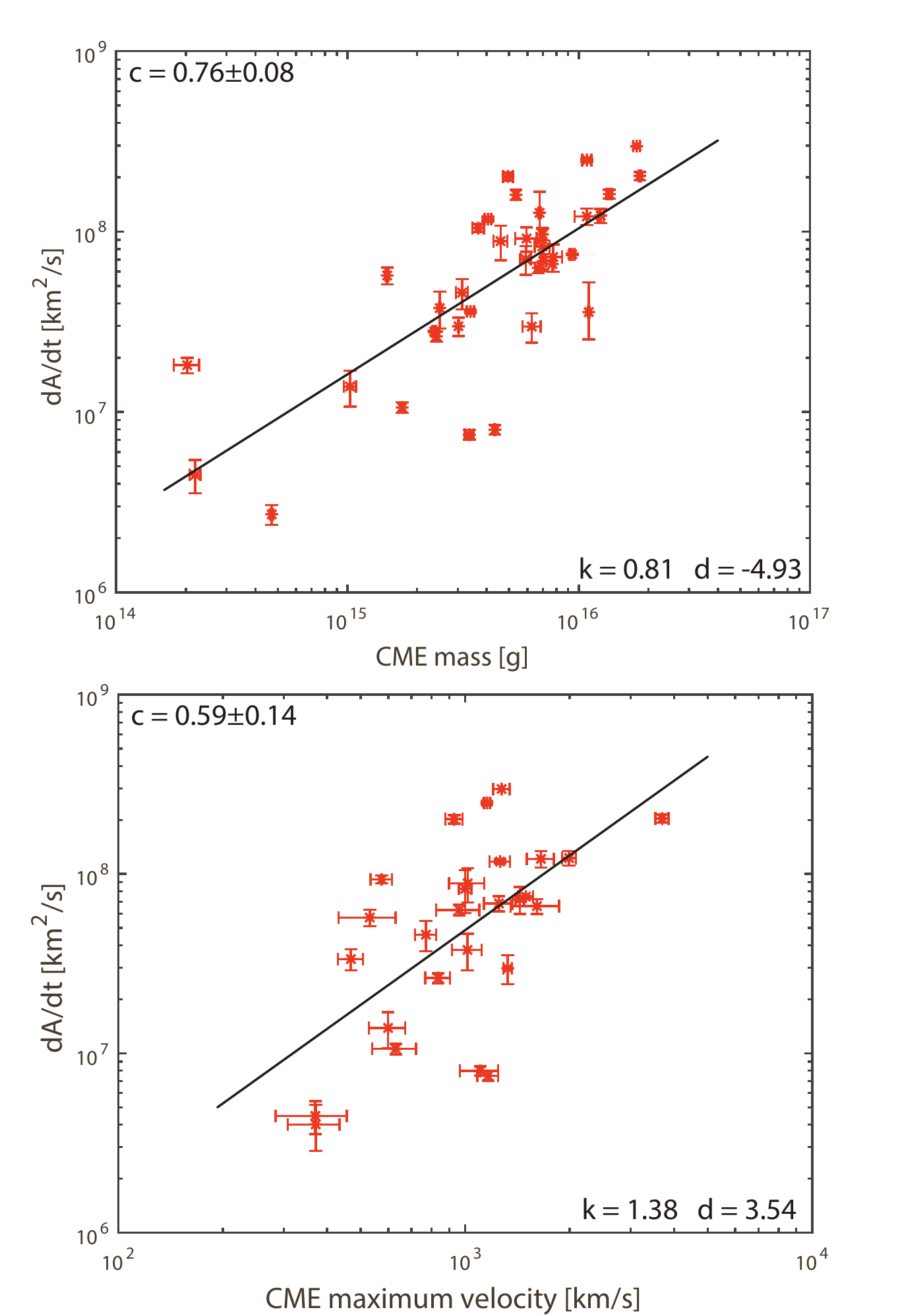}
\caption{Correlation of the area growth rate $dA/dt$ of the coronal dimming observed off-limb and the CME mass (top) and CME maximal speed (bottom). The black line represents the linear regression fit to all data points. The correlation coefficient is given in the top left corner of each panel. The parameters of the linear regression performed in log-log space are given in the bottom right corner of each panel.}
\label{fig:dA-CME}
\end{figure}

In addition, we also investigated the relation between the dynamic evolution of the dimming, as described by the peak of its area growth rate $dA/dt$, and the parameters of the associated CME. The correlation coefficient in logarithmic space are $c$ = $0.76\pm 0.08$ and $c$ = $0.59\pm 0.14$, respectively, indicating a close connection between the dimming growth rate with the mass and speed of the associated CME.

\begin{figure}[H]
\centering
\includegraphics[width=0.6\columnwidth]{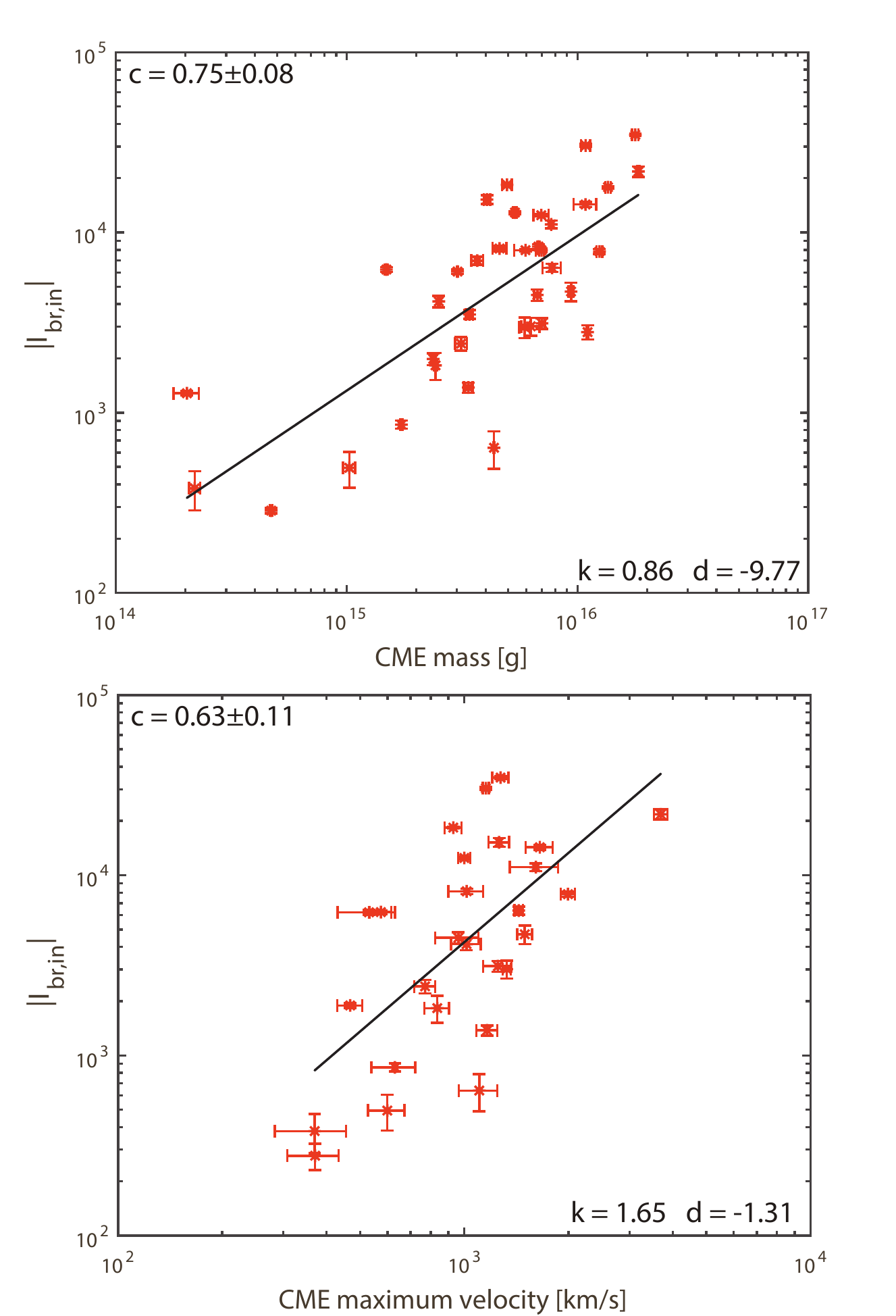}
\caption{Absolute values of the total instantaneous dimming brightness $|I_{br,in}|$ against the mass (top panel) and the maximum velocity (bottom panel) of the corresponding CME in logarithmic space. The brightness is calculated from LBR data. The black line represents the linear regression fit to all data points. The correlation coefficients are given in the top left corner of each panel. The parameters of the linear regression performed in log-log space are given in the bottom right corner of each panel.}
\label{fig:brightness-speed}
\end{figure}

\begin{figure}[H]
\centering
\includegraphics[width=0.6\columnwidth]{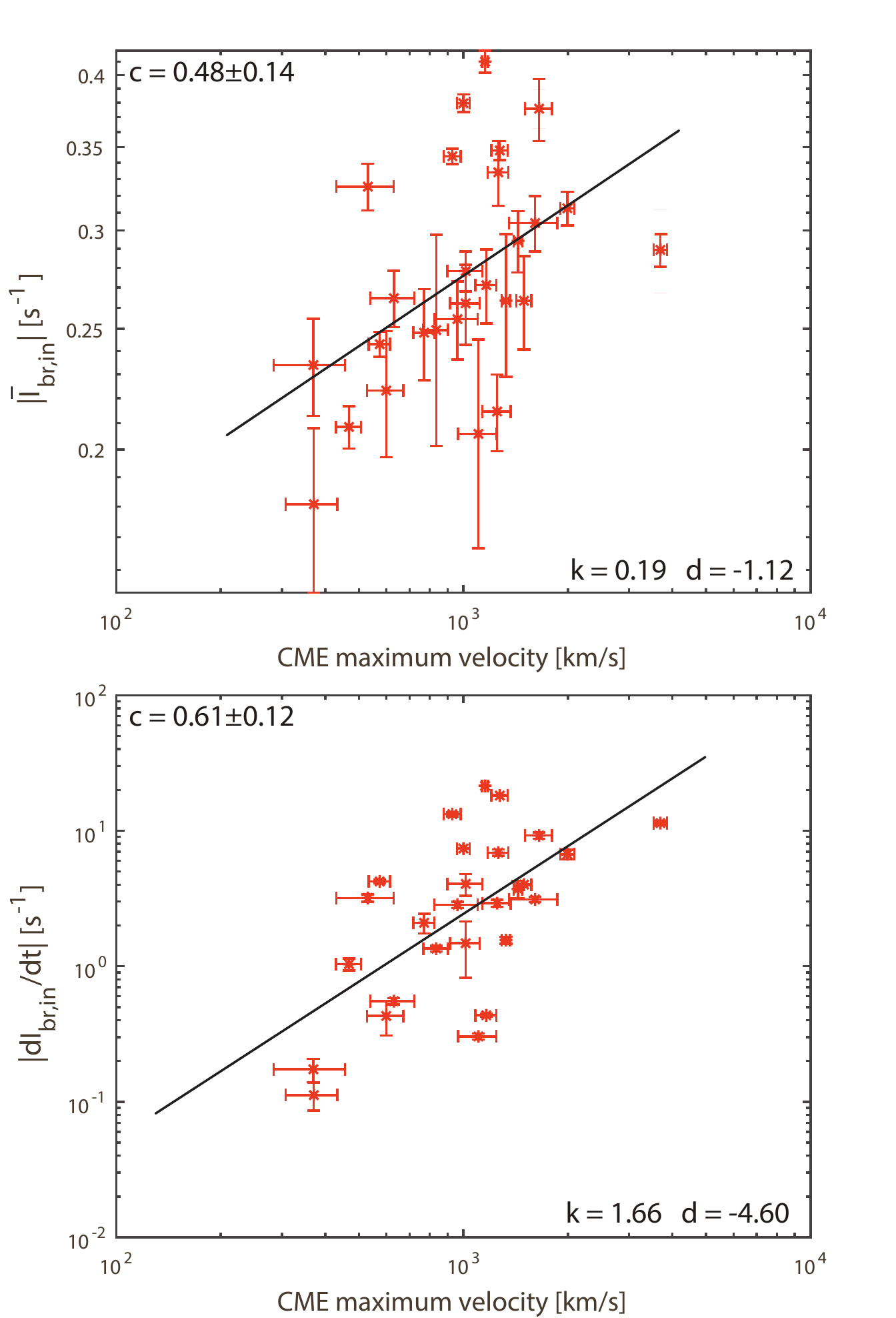}
\caption{Mean instantaneous dimming brightness $|\overline{I}_{br,in}|$(top panel) and instantaneous brightness change rate $|dI_{br,in}/dt|$ (bottom panel), calculated from LBR maps, against the maximum velocity of the associated CME in logarithmic space. All parameters are presented by their absolute values. The black line represents the linear regression fit to all data points. The correlation coefficients are given in the top left corner of each panel. The parameters of the linear regression performed in log-log space are given in the bottom right corner of each panel.}
\label{fig:mean_br}
\end{figure}

Figure \ref{fig:brightness-speed} shows the correlation plots of the absolute values of total instantaneous dimming brightness calculated from LBR data against the CME mass and maximal speed (bottom). The correlation coefficients are $c$ = $0.75\pm 0.08$ and $c$ = $0.63\pm 0.11$, respectively, indicating a strong correlation. In addition, in Figure \ref{fig:mean_br} we show the correlation between the mean instantaneous brightness $|\overline{I}_{br,in}|$ and the CME speed ($c$ = $0.48\pm 0.14$).

We also obtained a strong correlation between the dimming brightness change rate and the maximal speed of the CME: the faster the total dimming brightness is changing, the higher is the velocity of the associated CME. Figure~\ref{fig:mean_br} shows the absolute values of the instantaneous brightness change rate calculated from LBR data against the CME maximal speed. The correlation coefficient is $c$ = $0.61\pm 0.12$. 
\section{Summary and Discussion}
We have developed a robust automated algorithm for the segmentation of coronal dimmings above the solar limb in STEREO/EUVI 195 {\AA} images. The method is based on the combination of base-difference and logarithmic base-ratio data using a region-growing algorithm. This segmentation algorithm was applied to 43 off-limb coronal dimming events, and characteristic parameters describing their dynamic evolution, size, and brightness were derived. The events under study occured between May 2010 and September 2012, where the two STEREO satellites were close to quadrature position with respect to the Sun-Earth line. This unique setting allows us to derive the properties of the coronal dimmings above the solar limb in the STEREO/EUVI data, and for the first time to compare them with parameters of the same dimmings observed against the solar disk by SDO/AIA as well as with the associated CME kinematics and mass derived above the limb (minimizing projection effects) in \cite{dissauer2018statistics,dissauer2019statistics}. This approach provides us with important insight into the different projections of the coronal dimmings, and how they relate to the associated CME properties.

The off-limb dimming observations give us a line-of-sight integration across the CME from a lateral perspective. Thus, they provide us with a good view of its lateral expansion but the radial view is limited in height by the FOV of the EUVI imagers. The on-disk dimming observations are an integration of emission along the line-of-sight of the CME propagation, and thus correspond to a projection of the cross-section of the CME from a top view. However, they are affected by the “background” emission of the lower atmosphere layers, which makes it difficult to extract faint dimming regions and their effects higher up in the corona. On the other hand, on-disk dimmings provide us with the earliest insight into Earth-directed CMEs in observations from the Sun-Earth line, before the CME front reaches the FOV of white-light coronagraphs. Thus, comparing the properties of coronal dimmings observed simultaneously from different views in their on-disk/off-limb projections as well as their relation to the CME velocity and mass, allows us to assess the potential of using coronal dimmings for early CME characterization from satellites located either at L1 or L5.

\quad

Our main findings are as follows:
\begin{enumerate}

\item The derived off-limb dimming areas range from $7.01\times10^9$ km$^2$ to $5.48\times10^{11}$ km$^2$ (Fig.~\ref{fig:hist_parameters}). For the overlapping 39 events, which were studied also in \cite{dissauer2018statistics, dissauer2019statistics}, the mean value is $1.24\pm1.23\times10^{11}$ km$^2$, while the on-disk observations give $3.51\pm0.71\times10^{10}$ km$^2$. Only for 8 events the dimming area observed against the disk shows larger values than for the off-limb observations (see Fig.~\ref{fig:area-on-off}), with a correlation coefficient of $c = 0.63\pm 0.10$.

\item We compared the total dimming brightness calculated from LBR and BD data with the corresponding brightness extracted against the disk (Fig.~\ref{total-on-off}). The correlation coefficients are $c$ = $0.60\pm 0.14$ and $c$ = $0.77\pm 0.09$, respectively, and the slopes derived from the fits in logarithmic space are close to 1. This means that the entire decrease in the intensity of the dimming region observed on-disk and off-limb are linearly related to each other (with respect to the logarithmic space), but in most cases the on-disk dimming intensity is smaller (darker).

\item We also checked the duration of the impulsive dimming phase of 43 events obtained from off-limb observations (see panel (d) in Fig. \ref{fig:hist_parameters}). Although the time cadence of the STEREO data is much lower than the cadence of SDO/AIA, the duration parameter both for on-disk and off-limb observations is on average \mbox{$\sim$ 60 - 70 minutes}. 

\item The dimming recovery time $t_{rec}$ varies from 0.7 hrs to 10.9 hrs (Fig.~\ref{fig:time_rec}). The mean value is $4.6\pm2.8$ hrs, which supports the results by \cite{reinard2008coronal}, where the mean of the recovery time $4.8\pm0.3$ hrs was reported as well as with \cite{krista2017statistical} who found that the time from the area maximum to when dimming fully disappears is on average \mbox{$\sim6$ hrs}.

\item The CME mass shows the strongest correlation with the parameters, reflecting the total extent of the dimming, i.e. its area and total brightness: $c$ = $0.82\pm 0.06$ and $c$ = $0.75\pm 0.08$, respectively (see Fig.~\ref{fig:area_mass} and Fig.~\ref{fig:brightness-speed}). This result demonstrates that off-limb observations are able to provide a more accurate estimation of the CME mass.

\item The maximal CME speed correlates with the parameters describing the dynamics of the coronal dimmings: $c$ = $0.59\pm 0.14$ for the correlation with the area growth rate $dA/dt$ (see Fig.~\ref{fig:dA-CME}), $c$ = $0.61\pm 0.12$ for the correlation with the instantaneous brightness change rate, derived from LBR data (Fig.~\ref{fig:mean_br}). In addition, there is also a moderate correlation ($c$ = $0.48\pm 0.14$) of the CME maximal velocity and the mean brightness of the dimming (see also Fig.~\ref{fig:mean_br}).

\end{enumerate}   
 
 Figure~\ref{fig:area-on-off} shows that for the events under study, off-limb dimmings tend to be larger than their corresponding on-disk counterparts. Due to the line-of-sight integration, the density depletion higher up in the corona may not be detected for on-disk observations due to brighter lower-lying regions. Base-ratio images, showing relative changes in intensity, allow us to detect these regions from the off-limb perspective. At the same time, the area parameters show a strong correlation ($c\sim$ 0.63), which may indicate that mostly large dimmings are also detected in the higher corona.
 
Although typically the area of the dimming observed off-limb is larger than on-disk, the absolute total brightness obtained from off-limb base-difference data is lower than the one obtained from the on-disk data (see Fig.~\ref{total-on-off}). Dimming regions detected higher up in the corona, show less significant intensity decrease compared to dimming regions detected close to the surface where the density of the corona is higher. The biggest area contribution of dimming regions detected off-limb results from these regions higher up, but they do not contribute much to the overall intensity decrease of the dimming region. 
 
 We found a very strong correlation between the CME mass and the dimming area, which confirms the relation of coronal dimmings and CME physical properties for the off-limb viewpoint. This is similar to the previous findings from the on-disk study \citep{dissauer2019statistics}. More massive CMEs evacuate more plasma and create larger regions of EUV density depletion, which we detect from different lines of sight (see Fig.~\ref{fig:area_mass}). The correlation coefficient revealed an even higher value than for on-disk results ($c$ = $0.82\pm 0.06$ for off-limb against $c$ = $0.69\pm 0.10$ for on-disk). This may be related to the fact that the CME parameters were also extracted from STEREO data. Thus, the projection effects may cause similar uncertainties for the dimming quantities and the CME parameters derived.

The maximal area of the dimming also demonstrates a high correlation coefficient with the speed of the CME ($c=0.67 \pm 0.10$ in logarithmic scale). This relationship supports the finding that faster CMEs are usually also more massive \citep{mason2016relationship,aschwanden2016global,dissauer2019statistics} and therefore are also associated with a larger coronal dimming region. The correlation of the area growth rate with the CME speed and mass ($c\sim$ 0.6-0.7) supports this statement (Fig.~\ref{fig:dA-CME}).

  According to \citet{mason2016relationship}, CMEs that are associated with dark and large dimmings, tend to be more massive. The analysis of the dimmings on-disk by \cite{dissauer2019statistics} supports this statement: the correlation coefficient is $c\sim$ 0.6 for the correlation between the dimming total brightness and the CME mass. In the off-limb case the total dimming brightness $|I_{br,in}|$ reveals an even stronger correlation with the CME mass ($c\sim$ 0.75). Thus, dimmings can provide a measure of the amount of plasma evacuation in the CME: darker dimmings indicate a larger density depletion.

Furthermore, we found a moderate correlation ($c$ = $0.48\pm 0.14$) of the mean instantaneous brightness $|\overline{I}_{br,in}|$ with the CME speed (Fig.~\ref{fig:mean_br}). As the coronal plasma density strongly decreases with height, this correlation may indicate that faster CMEs tend to develop lower in the corona. This is consistent with studies of the CME source region characteristics and the fact that low in the corona also the magnetic field and thus the driving Lorentz force is stronger \citep{vrvsnak2007acceleration,bein2011impulsive}. The same correlation from the on-disk observations reveals \mbox{$c$ = $0.68\pm 0.08$}.

The dimming recovery times are suggestive of a bimodal distribution, with one group of $t_{rec}$ values smaller than 6 hrs and another group with values of $t_{rec}$ $>$ 6 hrs. This may indicate that there exist two different classes of dimmings as concerns the replenishment properties of the corona in the aftermath of a CME eruption. We also note that for \mbox{7 events} in our sample, the recovery time defined by the decrease of the instantaneous dimming area to 50\% of its peak value was not reached within the studied time range of 12 hrs, and may thus be even longer. These findings can be set in context with the detailed case studies of the plasma properties of several coronal dimming events using DEM analysis in \cite{2018ApJ...857...62V}. These authors found that in the localized core dimming regions, the density drops sharply within about 30 min and stays at these low levels for $>$ 10 hrs, whereas in the secondary dimming regions the density drop is more gradual, and the corona starts with the recovery and replenishment within 1-2 hrs. This finding was interpreted as evidence that the core dimmings are a signature of a flux rope still connected to the Sun, thus preventing the refill of coronal plasma in these regions. Further studies are needed and planned to obtain a better picture of the processes of the coronal replenishment and recovery after the wake of a CME, as is most prominently evidenced by the evolution of its dimming properties.

\section{Conclusion}
We performed a statistical analysis of 43 coronal dimming events that occurred during the time range between May 2010 and September 2012 that were observed simultaneously on-disk by SDO and off-limb by the STEREO satellites. The unique location of these satellites allowed us for the first time to look into the connection of the coronal dimmings and their associated Earth-directed CMEs statistically and to compare the results obtained from multi-viewpoint observations. Based on regression analysis, we confirm the relation of on-disk coronal dimmings and CME parameters presented in \cite{dissauer2019statistics} also for the off-limb viewpoint, for certain parameters providing even higher correlation coefficients than with the reported dimmings observed from on-disk by SDO AIA. Parameters describing the total dimming extent, i.e. area and the total brightness strongly correlate with the CME mass ($c\sim$ 0.7-0.8). The derivative of these parameters, i.e. the area growth rate and the brightness change rate show a high correlation with the CME speed (\mbox{$c\sim$ 0.6}), indicating the close relation between the CME and dimming dynamics.

In our observation range chosen (May 2010 - September 2012), the STEREO-A and -B satellites were located close to an L5/L4 configuration. The results from our study have therefore also relevant implications for the planned future L5 space weather mission, where Earth-directed CMEs, which are the most geo-effective, will be observed off-limb. The observations of coronal dimmings by solar EUV imagers  may help us to 
obtain a better characterization of Earth-directed CMEs, which is relevant for space weather applications. The distinct statistical relations derived between dimming parameters and decisive CME quantities for the different scenarios of L1 and L5 satellite locations (with correlation coefficients up to 0.8), provide a profound basis to improve early calculations of speed and mass of Earth-directed CMEs by including also the information of coronal dimmings.

\section{Acknowledgements}
The authors wish to thank the referee for his constructive and helpful comments that improved the manuscript. K.D. and A.M.V. acknowledge funding by the Austrian Space Applications Programme of the Austrian Research Promotion Agency FFG, BMVIT: projects ASAP-11 4900217, ASAP-14 865972, as well as the Austrian Science Fund (FWF):  projects P24092-N16,  P27292-N20. The STEREO/SECCHI data are produced by an international consortium of the Naval Research Laboratory (USA), Lockheed Martin Solar and Astrophysics Lab (USA), NASA Goddard Space Flight Center (USA), Rutherford Appleton Laboratory (UK), University of Birmingham (UK), Max-Planck-Institut f{\"u}r Sonnenforschung (Germany), Centre Spatiale de Li{\`e}ge (Belgium), Institut d'Optique Th{\'e}orique et Appliqu{\'e}e (France), and Institut d'Astrophysique Spatiale (France).

\bibliographystyle{apj}
\bibliography{main}

\end{document}